\documentclass[fleqn,usenatbib]{mnras}

\usepackage[T1]{fontenc} \usepackage{ae} \usepackage{aecompl}

\usepackage{amsmath}            
\usepackage{amssymb}            
\usepackage{graphicx}           
\usepackage[utf8]{inputenc}     

\usepackage{siunitx}            
\usepackage{newtxtext, newtxmath}

\defcitealias{Glatzle2019}{GCG19}
\defcitealias{draineCollisionalChargingInterstellar1987}{DS87}
\defcitealias{Weingartner2001b}{WD01}
\defcitealias{Weingartner2006}{WDB06}

\newcommand{\mrm}[1]{\mathrm{#1}}
\newcommand{\crash}{\textsc{crash}}
\newcommand{\diff}[1][\empty]{\ifx\empty#1
                \ensuremath{\mathop{}\!\mathrm{d}}
        \else
                \ensuremath{\mathop{}\!\mathrm{d}^#1}
        \fi}
\newcommand{\tder}[3][\empty]{\ifx\empty#1
                \ensuremath{\frac{\diff #2}{\diff #3}}
	\else
		\ensuremath{\frac{\diff[#1] #2}{\diff #3^{#1}}}
	\fi}

\sffamily 

\newcommand{\HII}{{\text{H}\ensuremath{^+}}}
\newcommand{\HeI}{{\text{He}\ensuremath{^0}}}
\newcommand{\HeII}{{\text{He}\ensuremath{^+}}}
\newcommand{\HeIII}{{\text{He}\ensuremath{^{++}}}}
\newcommand{\HIIt}{{\text{H}\,\textsc{ii}}}

\title[Grain charging in dustyCRASH]{Radiative transfer of ionizing radiation
  through gas and dust: grain charging in star forming regions}

\author[M. Glatzle et al.]{
Martin Glatzle,$^{1,2,6}$\thanks{E-mail: martin.glatzle@gmail.com}
Luca Graziani$^{3,4,5}$\thanks{E-mail: luca.graziani@uniroma1.it} and Benedetta Ciardi$^{1}$
\\
$^{1}$Max Planck Institut f\"ur Astrophysik, Karl-Schwarzschild-Str. 1,
85748 Garching, Germany\\
$^{2}$Physik-Department, Technische Universit\"at M\"unchen,
James-Franck-Str. 1,
85748 Garching, Germany\\
$^{3}$Dipartimento di Fisica, Sapienza, Universit$\grave{a}$ di Roma, Piazzale Aldo Moro 5, 00185, Roma, Italy \\
$^{4}$INFN, Sezione di Roma I, P.le Aldo Moro 2, 00185 Roma, Italy\\
$^{5}$INAF/Osservatorio Astrofisico di Arcetri, Largo E. Femi 5, 50125 Firenze, Italy \\
$^{6}$BooleWorks GmbH, Radlkoferstraße 2, 81373 München, Germany}

\date{Accepted XXX. Received YYY; in original form ZZZ}

\pubyear{2021}

\begin{document}
\label{firstpage}
\pagerange{\pageref{firstpage}--\pageref{lastpage}}
\maketitle

\begin{abstract}
  The presence of charged dust grains is known to have a profound impact on the
  physical evolution of the multiphase interstellar medium (ISM). Despite its
  importance, this process is still poorly explored in numerical simulations
  due to its complex physics and the tight dependence on the environment. Here
  we introduce a novel implementation of grain charging in the cosmological
  radiative transfer code \crash. We first benchmark the code predictions on a
  series of idealized dusty \HIIt\ regions created by a single star, in order
  to assess the impact of grain properties on the resulting spatial
  distribution of charges. Second, we perform a realistic radiative transfer
  simulation of a star forming region extracted from a dusty galaxy evolving in
  the Epoch of Reionization. We find that $\sim 13$\% of the total dust mass
  gets negatively charged, mainly silicate and graphite grains of radius
  \SI{e-3}{\micro m}. A complex spatial distribution of grain charges is also
  found, primarily depending on the exposure to stellar radiation and strongly
  varying along different lines of sight, as a result of radiative transfer
  effects. We finally assess the impact of grain properties (both chemical
  composition and size) on the resulting charge distribution. The new
  implementation described here will open up a wide range of possible studies
  investigating the physical evolution of the dusty ISM, nowadays accessible to
  observations of high and low redshift galaxies.
\end{abstract}




\section{Introduction}
\label{sec:introduction}
Recent high redshift observations (e.g. \citealt{knudsenMergerDustyGalaxy2017,
  tamuraDetectionFarinfraredIII2019, Hashimoto2019, bakxALMAUncoversII2020})
show that dusty galaxies are present in the young Universe, well before the end
of the epoch of reionization (EoR), at $z \gtrsim 6$
\citep{bowRebels2021}. Although detections at these redshifts are biased
towards bright objects, they indicate that the interstellar medium (ISM) of
many young galaxies is chemically evolved and could even contain a significant
amount of cosmic dust obscuring these objects in certain electromagnetic bands
\citep{2021Natur.597..489F}. In the Local Universe, where spatially resolved
observations are available \citep[e.g.][]{daviesDustPediaDefinitiveStudy2017,
  casasolaRadialDistributionDust2017, daviesDustPediaRelationshipsStars2019,
  lianouDustPropertiesStar2019}, cosmic dust is observed to co-evolve with its
environment \citep[e.g.][]{Draine2011}, so that its emission and absorption
features may be used as diagnostics for different phases of the ISM and
suggesting that it plays an important role in the life cycle of galaxies.

The evolution of a cosmic dust grain is driven by interactions with the
surrounding gas (e.g. collisions with ions and molecules), the ambient
radiation field (e.g. absorption, scattering, photo-electron emission), and
other grains (e.g. coagulation, shattering). These processes will generally
result in the grain obtaining a non-zero electric charge $eZ$\footnote{$e$ is
  the elementary charge.}, which in turn heavily affects subsequent
interactions. Grain charging also contributes to regulating the thermal status
of a plasma via photo-electric emission \citep[e.g.][]{Draine2011}. As
described in \cite{vanHoof2004}, small dust grains dominate the photo-electric
heating and cooling contribution of dust and could, in fact, be quite important
in certain environments.

In cold molecular clouds, likely shielded from UV radiation, grains can
interact with highly penetrating X-/cosmic rays and their secondary products,
inducing changes in grain charge. This process has a strong impact on grain
surface chemistry because it alters the rate of ion adsorption and the
efficiency of ice mantle formation
\citep[e.g.][]{ivlevInterstellarDustCharging2015}.

Where the density of dust is sufficiently high, as in proto-planetary disks or
dense cloud cores, the grain size/shape distribution evolves through
coagulation and fragmentation, processes that strongly depend on the charge of
interacting grains \citep[e.g.][]{dominikPhysicsDustCoagulation1997,
  akimkinInhibitedCoagulationMicronsize2020} and have profound implications for
the dust optical properties and thus for the extinction curve
\citep[e.g.][]{hirashitaDustGrowthInterstellar2012}. Recent, spatially resolved
ALMA\footnote{\url{https://www.almaobservatory.org}} observations of
proto-planetary discs \citep[e.g.][]{birnstielDiskSubstructuresHigh2018}, as
well as far infrared/sub-mm observations of variations in dust opacity along
ISM lines of sight containing dense clouds, provide evidence for such grain
size/shape evolution \citep[e.g.][]{kohlerDustCoagulationProcesses2012,
  tazakiUnveilingDustAggregate2019}.

In the diffuse, warmer ISM, grains are exposed to a wider photon energy range
and their interaction with the radiation field is complicated by the possible
presence of electrically charged polycyclic aromatic hydrocarbons
(PAHs)\footnote{PAH emission features are generally associated with atomic gas,
  while they are suppressed in ionized regions. The presence of PAHs in
  molecular gas is debated (see, e.g., \citealt[][\S~5.3]{Ferland2017} and
  \citealt{chastenetPolycyclicAromaticHydrocarbon2019}).}
\citep[e.g][]{siebenmorgenDustDiffuseInterstellar2014}. As the ambient medium
becomes hotter and ionized, grain destruction channels (e.g. sputtering) open
up. Their efficiency depends also on grain charge
\citep[e.g.][]{murgaShivaDustDestruction2019}.

Finally, in extreme environments illuminated by gamma-ray bursts or active
galactic nuclei, high energy photons could increase grain charges to excessive
values and induce Coulomb explosions
\citep[e.g.][]{waxmanDustSublimationGammaray2000,
  tazakiDustDestructionCharging2020}.

The depletion of certain atomic metals which enter the ISM in gaseous form to a
large fraction (e.g. Si), as observed in our Galaxy, is often interpreted as
evidence of grain growth, a process transferring metals from the gas phase onto
grains \citep[e.g.][]{whittetOxygenDepletionInterstellar2010,
  zhukovskaIronSilicateDust2018}. There is also a general consensus among
recent hydro-dynamical and semi-analytic models of galaxy formation
(e.g. \citealt{Mancini2015, mckinnonSimulatingDustContent2017,
  poppingDustContentGalaxies2017, gioanniniCosmicDustRate2017, Aoyama2018,
  grazianiAssemblyDustyGalaxies2020}) that some form of grain growth is
necessary to reproduce the total mass of dust in galaxies across cosmic
timescales, while its efficiency remains controversial
\citep{priestleyEfficiencyGrainGrowth2021}. When studying this process, grain
charge has to be considered, as the arrival rate of candidate ions for
accretion (e.g. Si$^+$) can be significantly modified by a non-zero charge
\citep[e.g.][]{zhukovskaIronSilicateDust2018}. Since grain growth taking place
in shielded, cold clouds seems difficult due to ice mantle formation
\citep{ceccarelliEvolutionGrainMantles2018}, photo-electric charging could play
a significant role in the growth process of all grains stable to radiation
exposure, taking place in a still unidentified phase of the ISM. Small grains,
which account for most dust surface area and are more likely to be negatively
charged compared to larger particles
\citep[e.g.][]{ibanez-mejiaDustChargeDistribution2019}, are primary candidates
for accreting positive ions
\citep[e.g.][]{weingartnerInterstellarDepletionVery1999} and thus accounting
for most of the ISM growth.

Charge is a parameter of fundamental importance to the dynamics of grains in a
plasma \citep[e.g.][]{melzerPhysicsDustyPlasmas2019}. This has been studied in
some detail in the interplanetary plasma
\citep[e.g.][]{mendisCosmicDustyPlasmas1994}, where direct observations and
even in-situ measurements are available. It likely also plays a crucial role in
stellar and galactic winds, which are often dusty
\citep[e.g.][]{veilleuxGalacticWinds2005}. Coulomb drag on charged grains, for
example, could significantly alter the dynamics of \HIIt\ regions
\citep{Akimkin2015, akimkinDustDynamicsEvolution2017} and
radiation-pressure-driven winds
\citep{hirashitaRadiationpressuredrivenDustTransport2019}.

We finally note that the interaction of photons with dust could be
significantly influenced by its charge. For example, absorption and scattering
cross sections of neutral and charged grains differ
\citep[e.g.][]{bohrenScatteringElectromagneticWaves1977,
  heinischMieScatteringCharged2012, kocifajScatteringElectromagneticWaves2012,
  kocifajOpticalPropertiesPolydispersion2012}. This could impact dust
extinction but also temperatures and emission spectra, especially of very small
grains. For low charge numbers ($\left|Z\right|\sim 1$), the change in the
absorption cross section of several PAHs has been investigated
\citep[e.g.][]{mallociOnlineDatabaseSpectral2007} and is accounted for in some
form by many dust models \citep[e.g.][]{Li2001, Draine2007,
  mulasModelingGalacticExtinction2013}. For other candidate components of
cosmic dust populations, such studies seem to be missing; the same applies to
radiative transfer models that self consistently account for this
effect. Furthermore, the photo-electric yield (\S~\ref{sec:photoCharging}) of
positively charged grains is reduced w.r.t. neutral grains; this inhibits
photo-electrons from carrying away energy absorbed by the grain and enhances
the probability of photo-destruction
\citep[e.g.][]{murgaShivaDustDestruction2019}.

Despite the relevance of the charging process, to our knowledge, no available
galaxy formation model explicitly accounts for it, since the physics regulating
the process of charging depends on a number of variables at the ISM scale,
poorly constrained or unresolved in numerical simulations: e.g. the free
electron temperature, the plasma ionization status and chemical composition, as
well as the presence of a UV and/or X-/cosmic ray flux. The relevant
environmental conditions are difficult to reproduce in sub-grid models. Also
note that all the above quantities can significantly vary both across galactic
environments \citep[e.g.][]{ibanez-mejiaDustChargeDistribution2019} and in
time, as a result of the progress of star formation. Dependencies on the solid
state properties of the grains (e.g. amorphous vs crystalline), on their shapes
(e.g. spherical vs aggregates, \citealt{maChargingAggregateGrains2013}), sizes
and chemical compositions (e.g. silicate vs carbonaceous dust) further
complicate the theoretical modelling of the process
(\citealt{draineCollisionalChargingInterstellar1987, Weingartner2001b,
  Weingartner2006}, hereafter
\citetalias{draineCollisionalChargingInterstellar1987, Weingartner2001b,
  Weingartner2006}).

Photo-ionization codes are a powerful tool to study grain charging and provide
insights for galaxy formation models, since they follow the radiation field and
gas ionization, which provides electrons to be accreted by grains. To our
knowledge, charging is only accounted for by \textsc{Cloudy}
\citep{Ferland2017, vanhoofCurrentFutureDevelopment2020} in 1D and
\textsc{mocassin} \citep{ercolanoMOCASSINFullyThreedimensional2003,
  Ercolano2005, ercolanoXRayEnabledMOCASSIN2008} in 3D, with both codes
providing steady state solutions, i.e. disregarding any temporal evolution.
Grain charging in \textsc{Cloudy} was first introduced by
\cite{baldwinPhysicalConditionsOrion1991}. Their approach consists of computing
a surface potential for each considered grain material (e.g. graphite,
silicate) at each point in space, neglecting grain size dependencies or
averaging them out. \textsc{mocassin} has adopted this approach and refined it
by explicitly considering the strong grain size dependence (see
\S~\ref{sec:dustPlasma}). In \cite{vanHoof2004} a major upgrade to the grain
model of \textsc{Cloudy} was performed. A Mie theory
\citep{mieBeitrageZurOptik1908, Hulst1957, Bohren1983} implementation, based on
\cite{hansenLightScatteringPlanetary1974}, was introduced in order to be able
to work with arbitrary size distributions of spherical grains of arbitrary
composition. In each bin of the size distribution chosen for each composition,
intrinsic (absorption and scattering opacities) and derived (temperature,
charge) physical grain properties are computed. The process of grain charging,
in particular, is implemented in the form of a ``hybrid'' model
\citep{vanHoof2004, abelIIRegionPDR2005}, which modifies the treatment
introduced in \citetalias{Weingartner2001b} with the aim of resolving the
charge distribution (see \S~\ref{sec:timeCharging}) of small grains (radius
$a<\SI{1}{nm}$), while computing an average charge for large grains.

In this work we present the implementation of the grain charging process in the
cosmological radiative transfer code \crash\ (\citealt{Ciardi2001, Maselli2003,
  Maselli2009, Pierleoni2009, Graziani2013, Hariharan2017, Graziani2018,
  Glatzle2019}, hereafter \citetalias{Glatzle2019}) and we discuss the spatial
charge distribution in typical \HIIt\ regions created by single stellar sources
(or populations including binaries) and in the cloudy environment typical of a
star forming region.

The paper is organized as follows: \S~\ref{sec:dustPlasma} briefly reviews
grain charging physics and the temporal evolution of the process. The numerical
implementation in \crash\ is presented in \S~\ref{sec:CRASH} and the results
are discussed in \S~\ref{sec:results}. More specifically,
\S~\ref{sec:black-body-source} discusses some ideal configurations of \HIIt\
regions around a single stellar source; \S~\ref{sec:BPASS-source} explores the
case of a young stellar population, including binary stars, embedded in a
previously dust enriched medium as it is generally found in cosmological
simulations accounting for dust; and finally, \S~\ref{sec:SF-region} explores a
realistic case of a star forming region extracted from a galaxy simulated by
the \texttt{dustyGadget} code. Technical details on recent improvements to our
framework can be found in App.~\ref{sec:new-dust-cross} describing new dust
cross sections and in App.~\ref{sec:dust-fix} providing details on algorithmic
improvements and better consistency with analytic solutions on a wide range of
gas densities polluted by dust. Finally, details on how electron yields are
computed can be found in App.~\ref{sec:photo-el-yields}, while further tests on
analytic solutions are briefly shown in App.~\ref{sec:comp-with-analyt}.

\section{Grain charging processes in a plasma}
\label{sec:dustPlasma}
In this section we summarize the principal grain charging processes active in
an astrophysical plasma pervaded by a radiation field. The plasma is described
by its number density $n_\mrm{gas}\,[\si{cm^{-3}}]$, its temperature
$T\,[\si{K}]$ and ionization state $x_i$ with $i\in \{\HII, \HeII, \HeIII\}$;
the cosmological abundance ratio is assumed for hydrogen and helium (i.e.
$f_\mrm{H}=0.92$ and $f_\mrm{He}=0.08$), while molecules and atomic metals are
not accounted for.

Hereafter, we follow the approach of \citetalias{Weingartner2001b}, who base
their prescription for collisional charging on
\citetalias{draineCollisionalChargingInterstellar1987}. In light of the
capabilities of our RT code \citep{Graziani2018}, we also account for the
extension of their photo-electric charging prescription to X-ray energies
described in \citetalias{Weingartner2006}. \citetalias{Weingartner2001b} and
\citetalias{Weingartner2006} use the framework of the Silicate-Graphite-PAH
model \citep[e.g.][]{Draine2007}, which, as in \citetalias{Glatzle2019}, we
will adopt where concrete values are required, while aiming to keep any
numerical implementation general.

\subsection{Collisional charging}
\label{sec:collCharging}
A dust grain will be subject to collisions with charged particles of any
species present in the plasma\footnote{Collisions with neutral particles are
  not of interest here.}. For large grains, the collisional cross section can
be approximated to first order, i.e. by assuming monopole-monopole interaction
between the particle and the grain, neglecting perturbations of the grain's
charge distribution\footnote{We note that here `` charge distribution'' refers
  to the physical distribution of charge inside a single grain, which is
  assumed to be spherically symmetric when undisturbed. It is different from
  both the distribution of charges in a grain population mentioned in the
  introduction and discussed in \S~\ref{sec:timeCharging} and the distribution
  of charge values in space as mentioned in our discussion of results.} caused
by the incoming particle. A proper treatment for small grains, on the other
hand, should account for the fact that the incoming charge could polarize the
grain, creating an "image potential"
\citep{jacksonClassicalElectrodynamics1962} and significantly enhancing the
cross section. We refer the reader to
\citetalias{draineCollisionalChargingInterstellar1987} for further details.

Free electrons are the most important collision partners due to their high
abundance and velocities. Upon collision with a grain, an electron may be
reflected or transmitted and escape to infinity. On the other hand, it may
stick to the grain with a probability given by the sticking coefficient
$s_\mrm{e}$, adding a negative charge and lowering $Z$. $s_\mrm{e}$ is highly
uncertain and it depends on grain properties (composition, $a$, $Z$) and the
electron's kinetic energy (the latter and the composition dependency are
neglected by \citetalias{Weingartner2001b}; see their \S~3.1). If the
electron's kinetic energy is high enough, the collision could free one (or
more) secondary electron(s), increasing $Z$ (see \citealt{Draine1979a}).

Collisions with ions present in the plasma (e.g. \HII) can also occur, but ions
are slower than electrons and will thus collide less frequently with dust
grains. If a collision takes place, the ion is most likely neutralized by
electrons from the grain\footnote{Neutralization of the incoming ions is likely
  to occur because the minimum energy required to remove an electron from a
  bulk solid (the so called work function, e.g.
  \citealt{willisOpticalPhotoemissionProperties1973}) is generally smaller than
  the ionization potential of individual atoms, although for grains finite size
  effects apply. \citetalias{Weingartner2001b} assume a constant neutralization
  coefficient $s_\mrm{n}=1$ for all ions.}, leading to an increase of $Z$. The
resulting atom may be released back to the gas phase or it may remain adsorbed
to the grain, the outcome being inconsequential to $Z$. Adsorption likely
occurs in partially ionized media, where grains carry negative charges (see the
introduction and, for an extended discussion,
\citealt{zhukovskaIronSilicateDust2018}).

Grain-grain collisions, finally, are rare in tenuous media and are thus
neglected in our discussion.

Given the above, the charging current experienced by a grain of some
composition due to thermal collisions with plasma species $i$ (excluding
secondary ionization in the case of electrons), can be expressed as follows:
\begin{equation}
  \label{eq:col-char-rate}
  J_i =
  q_i n_i s_i(a, Z)\sqrt{\frac{8kT}{\pi m_i}}\hat{J}(T,a,Z,q_i)\,,
\end{equation}
where $n_i$ is the number density of species $i$, $q_i$ its charge, $m_i$ its
mass and $s_i$ its sticking/neutralization coefficient. Expressions for
$\hat{J}$ can be found in
\citetalias{draineCollisionalChargingInterstellar1987}. The charging current
due to thermal electrons ejecting secondaries ($J_\mrm{sec, gas}$) has been
modelled by \citet[][\S~III~a)~ii)]{Draine1979a}.

\label{sec:collCharging}
\begin{figure}
  \centering
  \includegraphics[width=\linewidth]{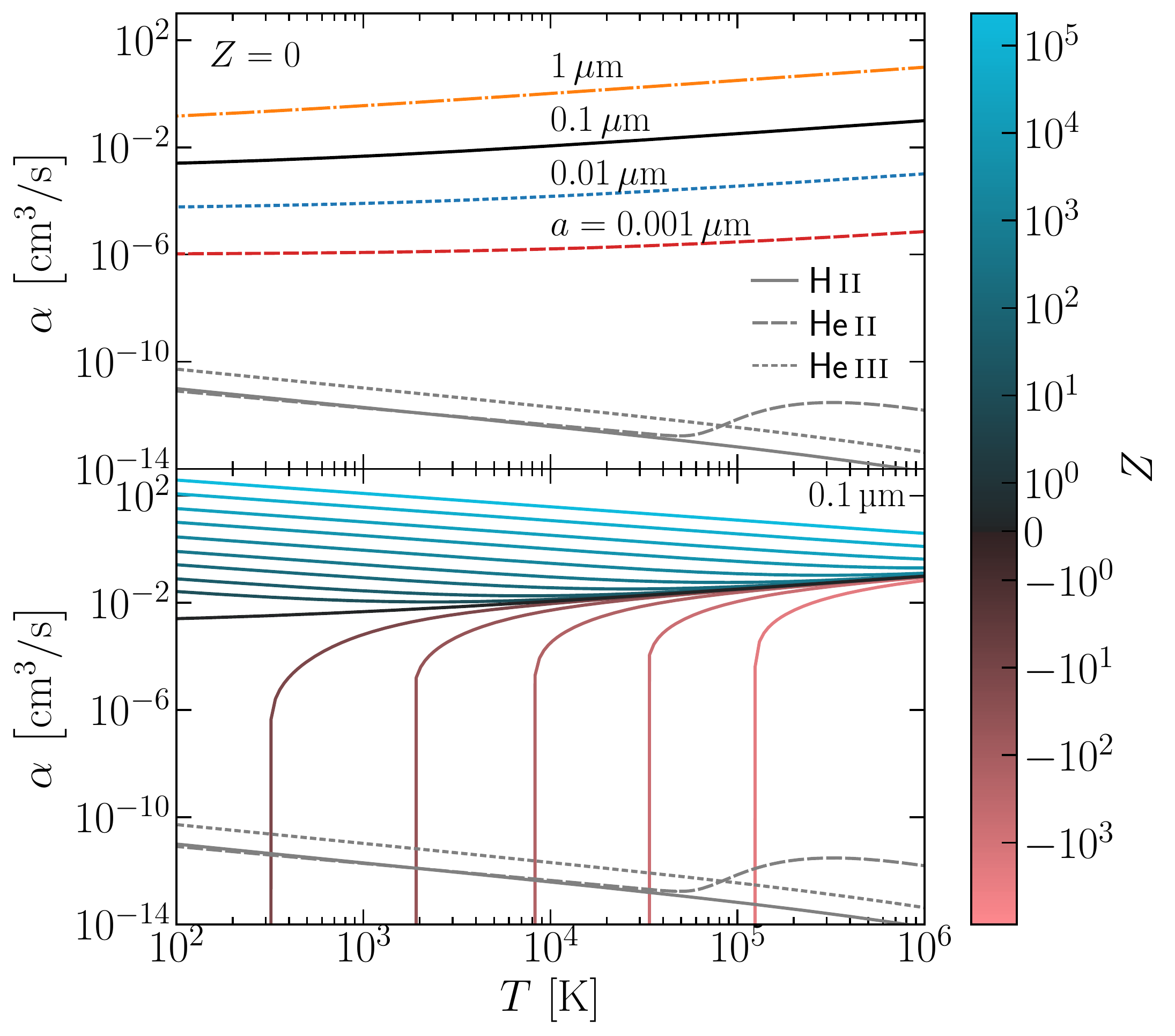}
  \caption{Electron charging rate coefficient as a function of plasma
    temperature for neutral grains of different sizes (top panel, indicated by
    the line style and colour) and for a grain of \SI{0.1}{\micro m} with
    different charge numbers (bottom panel, indicated by line colour as shown
    by the colour bar in symlog scale). Also shown for comparison are the
    hydrogen and helium recombination rate coefficients (grey lines).}
  \label{fig:recomb-rate-coeff-dust}
\end{figure}

In the top panel of Fig.~\ref{fig:recomb-rate-coeff-dust} we show the electron
charging rate coefficient $\alpha = J_\mrm{e}/(-e n_\mrm{e})$ as a function of
the plasma temperature for neutral grains of different sizes. Note that due to
the associated uncertainties, \citetalias{Weingartner2001b} provide identical
prescriptions for $J_\mrm{e}$ for both carbonaceous and silicate grains, even
though in reality there likely is a composition dependence. $\alpha$ varies
strongly with grain size, as expected due to the variation in the geometrical
cross section. Note, however, that despite their smaller sticking coefficient,
the rate coefficient for small grains is enhanced. One would expect a factor of
$\sim\num{e6}$ between the rates for the \SI{0.001}{\micro m} and the
\SI{1}{\micro m} grain, but at \SI{100}{K} it is only $\sim\num{e5}$ due to the
small grain effect discussed above.

In the bottom panel, we show $\alpha$ for $a=\SI{0.1}{\micro m}$ and different
charge numbers as indicated by the line colour. Hereafter, \SI{0.1}{\micro m}
is our reference grain size as it is typical of the diffuse ISM and it will be
represented by a solid black line in its neutral state. The lower limit on the
charge numbers shown is set by auto-ionization
(cf.~\citetalias[][Eq.~(24)]{Weingartner2001b}), while the upper limit is set
by Coulomb explosion, assuming a tensile strength
$S_\mrm{max}=\SI{e11}{dyn.cm^{-2}}$ for both grain materials
(cf.~\citetalias[][\S~10.2.1]{Weingartner2006}). The auto-ionization limit for
carbonaceous grains is slightly lower than for silicate grains, so it was
chosen here. Positive grain charges result in a higher charging rate due to
Coulomb focussing, which is more efficient for slower electrons (low
temperatures). For negative charges, the rate is zero at low $T$ until the
electrons have enough kinetic energy to overcome Coulomb repulsion. At this
point the rate rises very steeply, soon reaching values comparable to those of
the neutral case.

A similar analysis can be performed on the proton charging rate coefficient,
yielding similar results as per Eq.~\eqref{eq:col-char-rate}; it is thus not
repeated here for the sake of brevity.

\subsection{Photo-electric charging}
\label{sec:photoCharging}
\begin{figure}
  \centering
  \includegraphics[width=\linewidth]{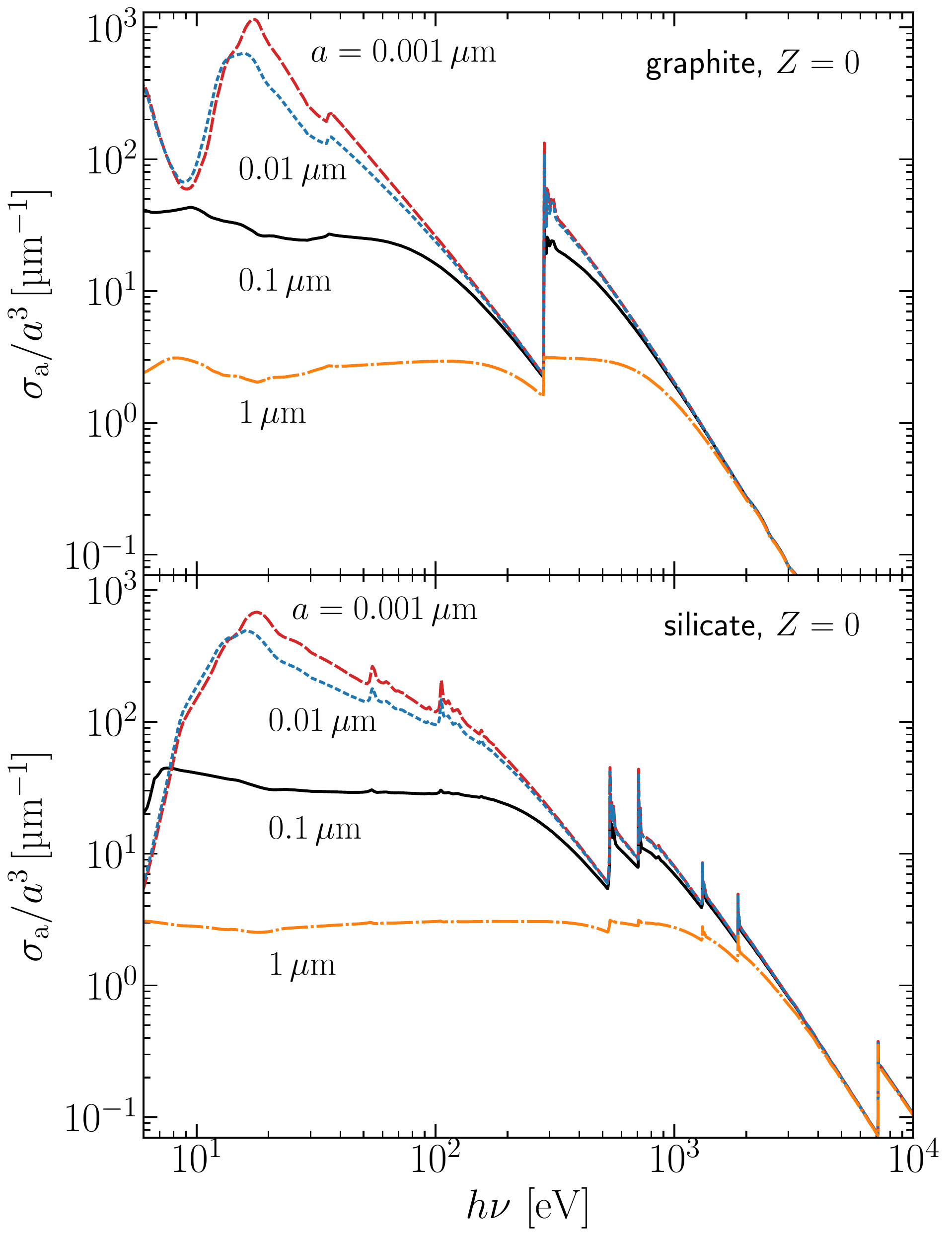}
  \caption{Absorption cross section normalized to volume for neutral graphite
    (top panel) and silicate (bottom panel) grains of different sizes
    (indicated by line style and colour) as function of photon energy.}
  \label{fig:sigma-d-sizes}
\end{figure}

An astrophysical plasma will generally be pervaded by electromagnetic
radiation, resulting in the absorption of photons by
grains. Fig.~\ref{fig:sigma-d-sizes} shows the volume-normalized absorption
cross section as a function of photon energy $h\nu$ for grains represented by
homogeneous spheres of graphite (top panel) and silicate (bottom panel) of
different radii, computed in the Silicate-Graphite-PAH framework as described
in App.~\ref{sec:new-dust-cross}. The computation assumes neutral grains, but
these values are also adopted for charged grains, since no prescription for the
computation of their cross sections is available\footnote{Cross sections for
  singly charged PAHs are provided (see \citealt{Li2001}), but they are
  identical to those of neutral PAHs for $h\nu \gtrsim \SI{4}{eV}$.}. For
$h\nu<\SI{200}{eV}$ and $a<\SI{0.1}{\micro m}$, different resonances are
present across grain compositions, while larger grains show a flat dependency
on $h\nu$, referred to as ``grey'' absorption. At X-ray energies, absorption
edges are visible since photons can interact with single constituent
atoms. Finally, note that the increase of $\sigma_\mrm{a}/a^3$ at
$h\nu < \SI{10}{eV}$ present for small graphite grains corresponds to the
\SI{2175}{\angstrom} extinction bump (cf.~\citealt[][Fig.~24.1]{Draine2011}).

\label{sec:photoCharging}

A photon absorbed by a dust grain may eject electrons from it, thus increasing
$Z$. The average number of electrons ejected following an absorption event
(with a fixed set of parameters) is referred to as photo-electric yield
$Y$. \citetalias{Weingartner2001b} have computed separate yields for
carbonaceous and silicate grains at low photon energies
($h \nu \leq \SI{20}{eV}$) accounting for the dependence on $a$ and $Z$. Note
that at these energies, electrons are ejected primarily from the band structure
of a grain.

\citetalias{Weingartner2006} updated the yield computation to treat additional
processes occurring for $\SI{20}{eV}< h\nu \lesssim\SI{10}{keV}$. High-energy
photons can eject electrons from the inner shells of a grain's atoms, possibly
inducing a cascade of Auger electrons if the first electron comes from a shell
below the highest occupied one. In analogy to the secondary ionization process
occurring in a gas photo-ionized by X-rays (see \citealt{Graziani2018}),
photo-electrons released with high energy can themselves excite secondary
electrons before leaving the grain. \citetalias{Weingartner2006} provide an
expression for each process and sum up all contributions to compute the total
yield $Y$:

\begin{equation}
  Y = Y_\mathrm{band} + Y_\mathrm{inner} + Y_\mathrm{A} + Y_\mathrm{sec}\,,
\end{equation}

where $Y_\mathrm{band}$ is the band structure yield, $Y_\mathrm{inner}$
accounts for inner atomic shells which are not part of the band structure,
$Y_\mathrm{A}$ is the associated Auger yield, and $Y_\mathrm{sec}$ denotes the
secondary yield, which is computed accounting for all of the three previous
yields.

\begin{figure}
  \centering
  \includegraphics[width=\linewidth]{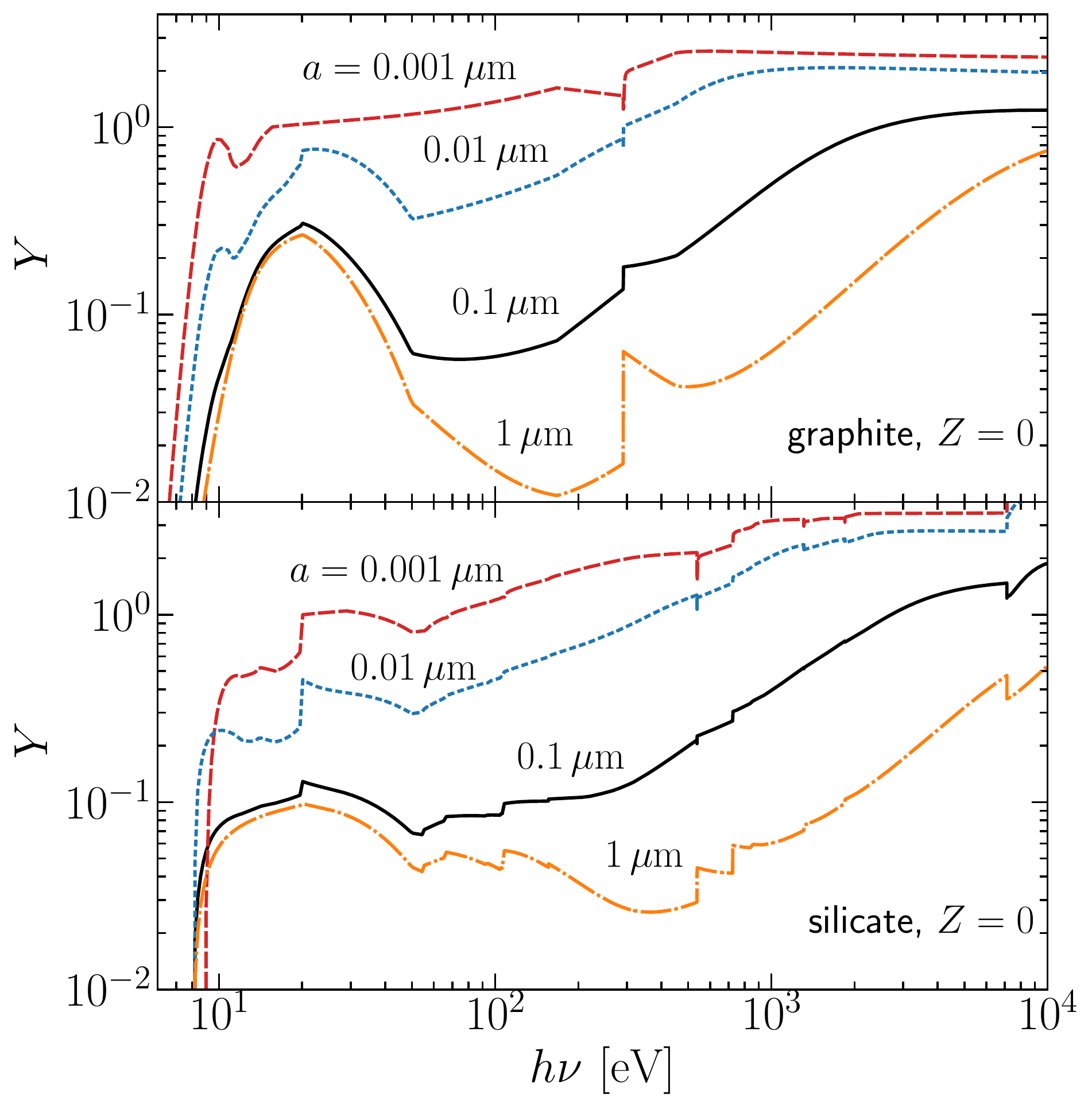}
  \caption{Photo-electron yield as a function of photon energy $h\nu$ for
    carbonaceous (upper panel) and silicate (lower panel) grains of various
    radii $a$ (indicated by line style and colour) with initial charge number
    $Z=0$.}
  \label{fig:pe-yield-sizes} \end{figure}

Fig.~\ref{fig:pe-yield-sizes} shows $Y$ for $Z=0$ as a function of $h\nu$, for
different grain radii and materials. The yields are computed using our
implementation following \citetalias{Weingartner2006} (see
App.~\ref{sec:photo-el-yields}) and are tabulated in order to be included in
our RT code (see \S~\ref{sec:chargedCRASH}).

Grains of our reference size \SI{0.1}{\micro\meter} (black lines) show an
increasing trend up to $h\nu\sim \SI{20}{eV}$ and a successive decrease until
$h\nu\sim \SI{50}{eV}$, where their yield starts to increase again up to X-ray
energies ($h\nu\sim\si{keV}$).  A strong dependence on the grain radius,
concerning both trends and absolute values, is evident by comparing with other
lines. Note, for example, that at $h\nu\sim \SI{300}{eV}$, the value of $Y$
(for both materials) changes by a factor of $\sim\num{100}$ as $a$ goes from
\SI{1}{\micro m} (big grains, orange line) to \SI{1}{nm} (very small grains,
red line). For very small grains, $Y$ rapidly reaches high values, meaning that
the photo-electric process is very efficient at all ionizing
energies. Comparing lines with the same colour across panels, one notices also
a strong dependence on chemical composition.

\begin{figure}
  \centering
  \includegraphics[width=\linewidth]{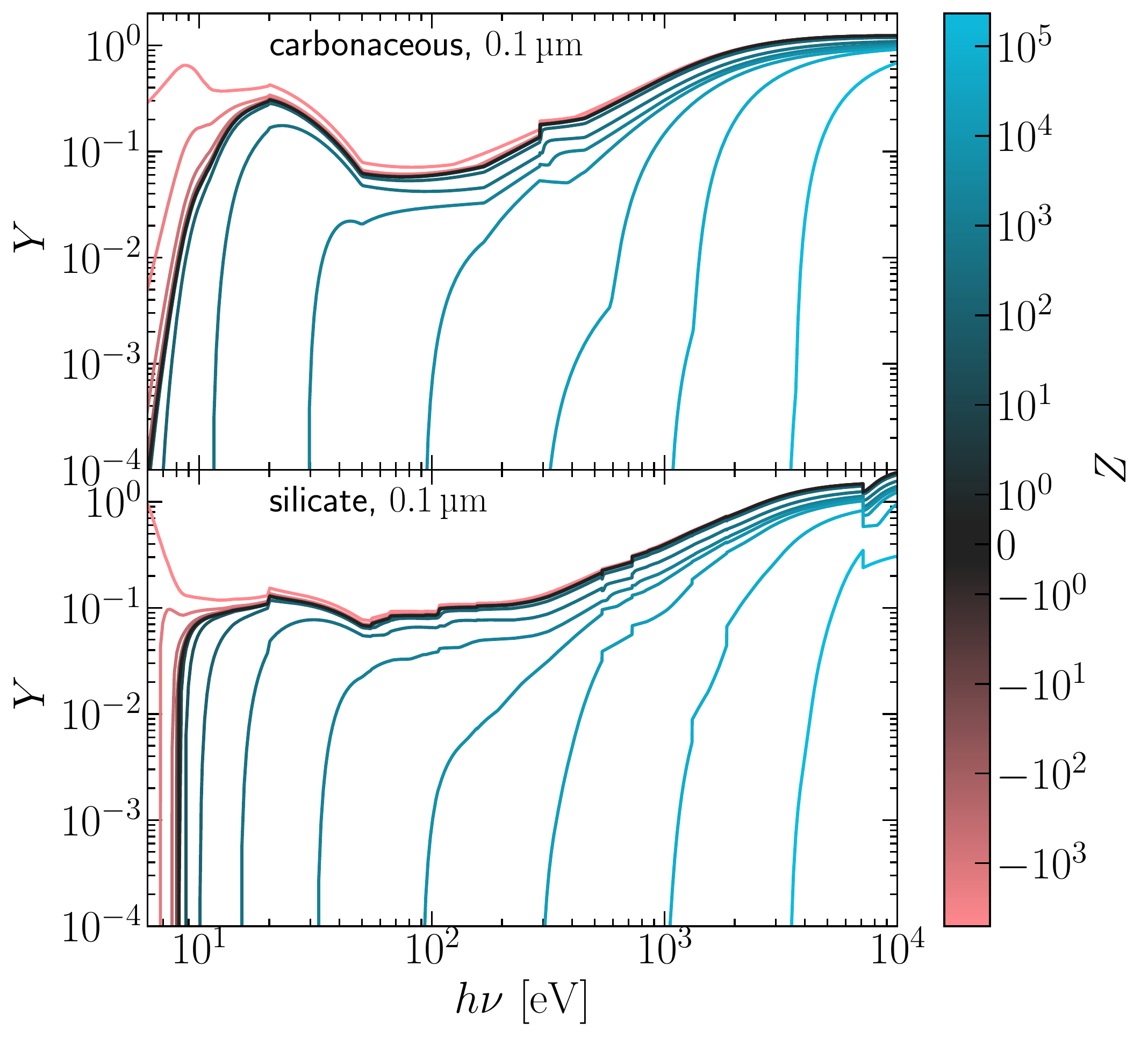}
  \caption{Photo-electric yield as a function of photon energy, $h\nu$, for
    carbonaceous (top panel) and silicate grains (bottom panel) of our
    reference grain radius $a= \SI{0.1}{\micro m}$. Line colours indicate
    different values of $Z$, as shown by the colour bar (symlog scale).}
  \label{fig:pe-yield-charge}
\end{figure}

The dependence of $Y$ on the initial $Z$ for our reference grain size is shown
in Fig.~\ref{fig:pe-yield-charge}. As in the bottom panel of
Fig. ~\ref{fig:recomb-rate-coeff-dust}, lines from red to light-blue indicate
the transition from the lowest to the highest possible
charge number. The Coulomb potential of positively charged grains, as expected,
rapidly suppresses the photo-electric efficiency as indicated by the sudden
decrease of $Y$ when $\log(Z) \geq 2$. For negative charge numbers, on the
other hand, the expulsion of an electron is more likely when the incoming
photons have energies $h\nu \leq \SI{20}{eV}$, while at higher energies
differences to the neutral case are weak. Again, a strong difference can be
seen between grain materials; note, for example, that at $h\nu < \SI{8}{eV}$
for very negatively charged silicate grains $Y\sim 1$, in opposition to the
neutral case.

In analogy to the collisional case, given a radiation field with specific
radiation energy density $u_\nu$, we can express the charging current
experienced by a grain of some composition due to photo-ejections as follows:

\begin{equation}
  J_\mathrm{pe}
  = e\int\diff{\nu}\frac{c u_\nu}{h\nu}\sigma_\mrm{a}(\nu,a)Y(\nu,a,Z)\,.
\end{equation}

The excess electrons attached to negatively charged grains are assumed to
occupy energy levels above the valence band by
\citetalias{Weingartner2001b}. As a result, photo-detachment of these electrons
has to be accounted for in the form of a separate current $J_\mrm{pd}$, which
is computed analogously to $J_\mrm{pe}$:

\begin{equation}
  J_\mathrm{pd} = e\int\diff{\nu}\frac{c u_\nu}{h\nu}\sigma_\mrm{pd}(\nu,a,Z)Y_\mrm{pd}\,,
\end{equation}

with the cross section $\sigma_\mrm{pd}$ described in
\citetalias[][\S~2.3.3]{Weingartner2001b} and $Y_\mrm{pd} = 1$\,.

\subsection{Time evolution of a grain's charge}
\label{sec:timeCharging}
Not all charging currents discussed in \S~\ref{sec:collCharging} and
\S~\ref{sec:photoCharging} are relevant under all circumstances. When
considering ionizing radiation, for example, $J_\mrm{pd}$ is subdominant to
$J_\mrm{pe}$. Where a strong radiation field is present (i.e. close to ionizing
sources), $J_\mrm{pe}$ will also dominate over collisional charging with ions
($J_\mrm{ion}$). Finally, in photo-ionized regions, where usually
$T\sim\SI{e4}{K}$, $J_\mrm{sec,gas}$ provides a second order contribution. For
these reasons, in the discussion presented here, $J_\mrm{pd}$ and
$J_\mrm{sec,gas}$ will be neglected. Moreover, the contribution of ions other
than protons to $J_\mrm{ion}$ will be neglected, i.e. we assume
$J_\mrm{ion} = J_\mrm{proton}$.

In analogy to the case of ions (e.g. \HeI, \HeII, \HeIII), for a given set of
conditions, a grain can be in a number of different charge states with varying
probability. The associated probability distribution is referred to as grain
charge distribution \citepalias[e.g.][Fig.~9]{Weingartner2001b}. The number of
available and typically occupied charge states, especially for large grains,
however, is many times higher than in the case of ions. For this reason, we
will use a simplified treatment here and consider the temporal evolution of
only a single charge\footnote{Note that this charge is not quantized. It can be
  thought of as an approximation to the average charge of a population of
  grains, but it will generally differ to some degree from the average of their
  real charge distribution.} for given grain properties, described by:
\begin{equation}
  \label{eq:timeCharge}
  e\tder{Z}{t} = J_\mrm{pe} + J_\mrm{ion} + J_\mrm{e}\,.
\end{equation}
In a steady state\footnote{The steady state solution corresponds to the
  ``equilibrium charge'' of \citetalias[][\S~7]{Weingartner2006}, which they
  find to be close to the average of the real charge distribution for large
  grains and even for small grains if charge numbers are high. The equilibrium
  charge is also what is computed in the approaches of \textsc{mocassin} and
  \textsc{Cloudy} (for large grains).}, the above equation becomes a root
finding problem in $Z$\footnote{So far we made an effort to clearly distinguish
  a grain's charge and its charge number; henceforth, for simplicity, $Z$ will
  refer to both as we see little potential for confusion.}:
\begin{equation}
  \label{eq:timeChargeSteadyState}
  J_\mrm{pe}(Z) + J_\mrm{ion}(Z)+ J_\mrm{e}(Z) = 0\,.
\end{equation}

\section{The cosmological radiative transfer code \crash}
\label{sec:CRASH}

\crash\ is a cosmological radiative transfer (RT) algorithm first introduced in
\cite{Ciardi2001} to simulate cosmic reionization around the first stars. The
RT is based on two common concepts: domain discretization and Monte Carlo (MC)
sampling of the radiation field using a long-characteristics ray tracing
scheme. Their combination allows to describe the radiation-matter interaction
occurring in each cubic cell of a Cartesian grid, once it is crossed by a ray
emitted from an arbitrary combination of sources. Both point-like sources, with
arbitrary spectra \citep{Maselli2009, Graziani2013}, or a diffuse background
field \citep{Maselli2003, grazianiUVBackgroundFluctuations2019} can be chosen,
depending on the astrophysical problem under investigation. \crash\ is usually
adopted to perform simulations of cosmological reionization of hydrogen and
helium \citep{Ciardi2003, ciardiSimulatingIntergalacticMedium2003, Ciardi2012,
  Eide2018a, eideLargescaleSimulationsHe2020} or to investigate specific
problems linked to the fluctuations of the UV background at fixed redshift
\citep{maselliRadiativeTransferEffects2005,
  grazianiUVBackgroundFluctuations2019}. Small scale cosmological environments,
such as the ones involved by quasar \HIIt\ regions
\citep{maselliProximityEffectHighredshift2004, Kakiichi2017, Graziani2018} or
Local Group-like volumes \citep{Graziani2015} have been also investigated,
proving the flexibility of the \crash\ algorithm in simulating a wide range of
cosmological environments.

As the general \crash\ algorithm is extensively described in a series of
technical papers (\citealt{Maselli2003, Maselli2009, Graziani2013,
  Graziani2018}; \citetalias{Glatzle2019}), here we provide the minimal outline
necessary to present how Eq.~(\ref{eq:timeChargeSteadyState}) is solved within
the time dependent RT scheme.

During the ray tracing cycle of \crash, each time $N_{\nu}$ photons
(distributed in a frequency bin around $\nu$) cross a cell along a cast path
$l$, their number is reduced by:
\begin{equation}
  \label{eq:photon-absorption-in-cell}
  \Delta N_\nu = N_\nu (1- \mathrm{e}^{-\tau_\nu})\,,
\end{equation}
where $\tau_\nu = \sum_i \tau_{i,\nu}$ is the total optical depth of the cell
at frequency $\nu$ and $i$ indicates the various species present in the
environment, either neutral atoms, ions or dust. The partial optical depth
contributed by each $i$ is given by
\begin{equation}
\tau_{i,\nu} = n_i\sigma_i l,
\end{equation}
where $n_i$ and $\sigma_i(\nu)$ are number density and absorption cross section
of the species $i$, respectively. The absorbed photons are then distributed
among $i$ as:
\begin{equation}
  \label{eq:ionization-number-in-cell}
  \Delta N_{i,\nu} = \frac{\tau_{i,\nu}}{\tau_\nu} \Delta N_\nu\,.
\end{equation}

This is used to estimate the time dependent photoionization rates for the
different gas species and the associated heating contributions. Taking into
account recombination, collisional ionization and various cooling processes,
the ionization state and temperature of the gas are then updated by evolving
the corresponding set of coupled, time-dependent differential equations (see
original papers for more details). In \citetalias{Glatzle2019} the energy
deposited in the dust component was simply neglected and the corresponding
photons removed from the total flux. In the next section we will detail how
they are used in this release of the code to compute the effects of grain
charging.

\subsection{Grain charging implementation}
\label{sec:chargedCRASH}
In this section we describe the numerical implementation of grain charging in
the latest release of \crash\ \citep{Graziani2018}, extended in
\citetalias{Glatzle2019} to account for dust absorption in the hydrogen
ionizing band $h\nu \in [\SI{13.6}{eV}, \SI{10}{keV}]$. The code version
adopted here features a significant number of improvements in both algorithmic
and engineering aspects, resulting, most significantly, in greatly improved
behaviour when simulating media with high dust density, as documented in
App.~\ref{sec:dust-fix}.

To account for grain charging, we first expanded the list of initial conditions
to include 3D maps of the charge at simulation start of each dust species $k$,
$Z_k(t_0)$. Furthermore, the DUST\_SPECIES \textsc{type} has been extended to
load tables of the electron charging rate coefficient as a function of plasma
temperature and grain charge $\alpha_k(T, Z)$, as well as the photo-electric
yield as a function of photon energy and grain charge
$Y_k(h\nu, Z)$\footnote{Note that the composition and size dependencies have
  been absorbed into $k$.}. The tables can be provided by any dust model,
keeping the implementation general, while the data-set of built-in models has
been expanded by including tables of the relevant quantities for the
Silicate-Graphite-PAH model, computed as discussed in
\S~\ref{sec:dustPlasma}. The DUST\_IN\_CELL \textsc{type} has been consistently
extended to store the charge of each species in the cell.

A fully self consistent treatment of dust charging (neglecting the grain charge
distribution; cf.~\S~\ref{sec:timeCharging}) in our time dependent scheme would
require extending the system of equations describing the gas component by a set
of equations (one per dust species) in the form of
Eq.~(\ref{eq:timeCharge}). Since grain charging timescales
($\sim e/(J_\mrm{pe}-J_\mrm{e})$, \citealt{draineElectricDipoleRadiation1998})
can be many orders of magnitude shorter than hydrogen and helium
ionization/recombination timescales
(cf.~Fig.~\ref{fig:recomb-rate-coeff-dust}), time integration of such a system
is extremely challenging. The timesteps adopted in \crash\ are adaptive, since
a cell is updated every time it is crossed by a photon packet, but they are on
the order of the gas recombination time scale and thus generally much longer
than the dust charging timescale. Reaching small enough timesteps would require
prohibitive numbers of photon packets or a complete restructuring of the
\crash\ algorithm to enable the use of implicit solvers, which are known to be
difficult to combine with MCRT schemes
\citep{noebauerMonteCarloRadiative2019}. We therefore take a different approach
and assume grain charges to always be in steady state at the timescales \crash\
resolves, i.e. we evolve the grain charge of each species in each cell from
steady state to steady state as the environment (radiation field, temperature,
ionization state) changes.

More precisely, every time a cell is crossed by a photon packet, we compute the
number of photons of each frequency absorbed by each dust species
$\Delta N_{k,\nu}$ using Eq.~(\ref{eq:ionization-number-in-cell}). Since it is
heavily affected by MC sampling noise, we accumulate this quantity over a
number $N_\mrm{d}$ of packet crossings
($\Delta N_{k,\nu}^\mrm{acc} = \sum_\mathrm{crossings} \Delta N_{k,\nu}$) to
obtain a robust estimate of its time average
$\Delta N_{k,\nu}^\mrm{acc}/\Delta t_\mrm{d}$, where $\Delta t_\mrm{d}$ is the
time since the last charge computation. We then use a root finder to solve
Eq.~(\ref{eq:timeChargeSteadyState}) for the new charge value $Z_k$,
approximating the photo-electric charging current:
\begin{equation}
  \label{eq:j-pe-crash}
  J_{\mrm{pe},k}
  = \frac{e}{\Delta t_\mrm{d}}\sum_\nu \Delta N_{k,\nu}^\mrm{acc}Y_k(h\nu,Z_k)\,,
\end{equation}
and the electron charging current:
\begin{equation}
  \label{eq:j-e-crash}
  J_{\mrm{e},k}
  = -e n_\mrm{e} \alpha_{\mrm{e},k}(T, Z_k)\,,
\end{equation}
where $Y_k$ and $\alpha_k$ are computed by interpolation on the pre-loaded
tables, and $n_\mrm{e}$ and $T$ take the current cell values.  $J_\mrm{ion}$ is
implemented analogously to $J_\mrm{e}$, replacing electron properties with the
corresponding proton properties. We verified that a value of
$N_\mrm{d} \sim 100$ is typically required to overcome MC noise\footnote{The
  total number of packets crossing a typical cell during a simulation is
  $\gg N_\mrm{d}$.}. Finally, note that while we account for the dust
contribution to $n_\mrm{e}$, we do not yet account for the dust contribution to
cooling and heating of the plasma.

A detailed validation of our approach in complex time dependent configurations
is still outstanding and requires convergence tests in $N_\mrm{d}$. This is
particularly important in the case of very small grains, as their charging
timescales could reach values comparable to those of the gas. The results
presented below are not affected by the above issue, since they correspond to
global equilibria of their respective configurations. Moreover, all results
presented in \S~\ref{sec:results} have been validated by convergence tests in
both photon sampling and ranges of assumed $N_\mrm{d}$.

In \citetalias{Glatzle2019}, our simulations included only one dust species
with properties corresponding to the average of a grain population. When
considering only absorption, this does not constitute an approximation
\citep{Steinacker2013}. However, since we now account also for charging, we
have to resolve a given grain population in composition and size, as these
strongly affect the charging process. A typical run of the updated code might
now include eight dust species (see below). While accurate in terms of
implementation, this is demanding in terms of memory requirements, which scale
with the product of the number of computational cells, dust species and
frequency bins (to compute $\Delta N_{k,\nu}^\mrm{acc}$). The memory cost is
especially high in cases investigating ISM environments, where the dust
generally pollutes a large part of the domain, but in realistic cosmological
simulations the dust-contaminated domain is usually restricted to the
environments of star-forming galaxies, significantly limiting the number of
enriched cells.

The new features documented above will be employed in the following to first
study a series of \HIIt\ region configurations. We consider eight dust species,
spanning the radii $a \in \{0.001, 0.01, 0.1, 1.0\}\,\si{\micro m}$ and the
compositions graphite and silicate in the Silicate-Graphite-PAH framework. We
note that we use carbonaceous charging properties ($\alpha$ and $Y$,
cf.~\S~\ref{sec:dustPlasma}) for the graphite grains, since no pure graphite
prescription is available\footnote{Note that while not fully self-consistent as
  small carbonaceous grains are tuned to the properties of PAHs, given the
  associated uncertainties this assumption is certainly acceptable.}.

\section{Results}
\label{sec:results}
In this section we investigate the grain charges attained in \HIIt\ regions
created by sources with different multi-frequency spectra. In all cases we
assume a realistic dust population consisting of the eight species mentioned
towards the end of \S~\ref{sec:chargedCRASH}. Since for the time being we do
not account for PAHs, we use the graphite/silicate size distribution from
\citet[][MRN hereafter]{Mathis1977} extended down to \SI{0.001}{\micro m} with
normalizations as given by \cite{Weingartner2001a} to determine their relative
abundances.

In detail, \S~\ref{sec:black-body-source} describes the spherically averaged
profiles of an ideal \HIIt\ region created by a black body source, while the
ideal case created by the spectrum of a young stellar population including
binaries is investigated in \S~\ref{sec:BPASS-source}. In both cases, we
consider only H-ionizing frequencies ($h\nu>\SI{13.6}{eV}$). Finally, the grain
charges obtained in a more realistic case are discussed in
\S~\ref{sec:SF-region}, where dust, gas and stellar source distributions are
motivated by a hydrodynamical simulation and the spectrum of each stellar
population is modelled using BPASS
\citep{stanwayReevaluatingOldStellar2018}. In this case, we extend the spectra
down to \SI{6}{eV}\footnote{It is important to remind here that as shown in
  Fig.~\ref{fig:pe-yield-sizes} and Fig.~\ref{fig:pe-yield-charge}, these
  photons, while not capable of ionizing hydrogen and helium, can interact with
  dust providing a non negligible contribution to grain charging.}.

\subsection{Ideal \HIIt\ region of a black body source}
\label{sec:black-body-source}
Here we adopt the basic simulation setup of the reference run in
\citetalias[][\S~5.1]{Glatzle2019} and thus only summarize its most important
properties. A source with a black body spectrum ($T_\mrm{bb}=\SI{e4}{K}$, and
emissivity $\dot{N}=\SI{e49}{s^{-1}}$) is placed in the corner of an
$(\SI{85}{pc})^3$ box with a constant gas density $n_\mrm{g}=\SI{1}{cm^{-3}}$
and cosmological abundances of H and He mapped onto a Cartesian grid with
$256^3$ cells. The medium is polluted with dust at a gas-to-dust ratio (GDR) of
\num{124}. Initially, gas and dust are neutral and the gas temperature is
$T=\SI{e2}{K}$. We run the simulation for \SI{10}{Myr}, ensuring that
equilibrium in temperature and ionization is reached, while using \num{2e9}
photon packets to obtain good MC convergence (to the per mill level in the
volume averaged gas ionization fractions).

\begin{figure}
  \centering
  \includegraphics[width=\linewidth]{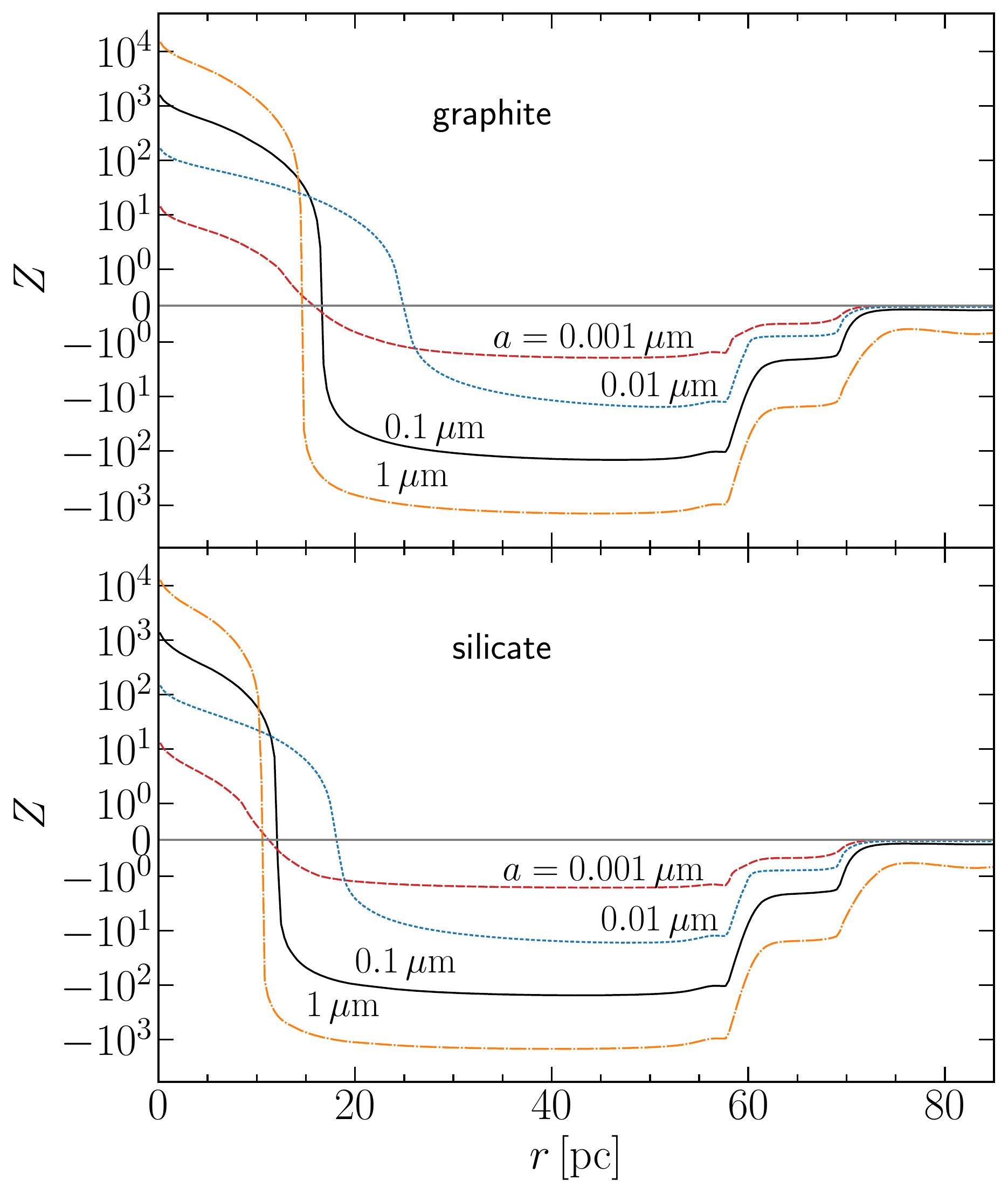}
  \caption{Spherically averaged grain charge at equilibrium as a function of
    distance $r$ from an ionizing black body source. The simulation includes
    eight species of grains with different sizes (line styles) and chemical
    compositions (panels). The horizontal line, corresponding to a neutral
    grain, is drawn to guide the eye. See text for details.}
  \label{fig:BB_Z_all}
\end{figure}

Fig.~\ref{fig:BB_Z_all} shows the spherically averaged profiles of the charges
of grains of different sizes (lines) and materials (panels) in the final
equilibrium configuration. Close to the source, all grains are positively
charged, but they attain negative charges as the photon flux becomes weaker
when moving away from the source. At the hydrogen ionization front
($\sim\SI{60}{pc}$), the number density of free electrons is drastically
reduced, resulting in grain charges becoming more positive again. In contrast
to the mono-frequency cases presented in App.~\ref{sec:comp-with-analyt},
however, the residual flux of high-energy photons beyond the ionization front
results in weak ionization of the gas and thus a non-zero number density of
free electrons, which keeps the charges negative instead of neutral outside the
front. Further comparisons with equivalent simulations neglecting the
contribution of $J_\mrm{ion}$ also confirm that the implementation of this
current is necessary when the flux of ionizing radiation is reduced
significantly, i.e. at $r > \SI{30}{pc}$. It is responsible, in particular, for
the value of the charge to progressively approach its neutral, initial value at
$r > \SI{70}{pc}$. Also note that in the external regions of an idealized
configuration like the one adopted here, the ionizing flux becomes so weak that
numerical artefacts can affect the stability of the values when computing the
spherical average for all quantities, including the predicted charge. This is
the case, for example, of \SI{1}{\micro m} grains (orange line), whose trend
starts to oscillate instead of becoming neutral at large distances. In the next
section this issue will be solved by adopting a more realistic source spectrum
extending to higher frequencies.

Finally note that even the highly idealized \HIIt\ region investigated here features a
complex charge structure, predicting spatially strongly varying conditions for
grain evolution.

\subsection{Ideal \HIIt\ region created by a young stellar population}
\label{sec:BPASS-source}
In this section we investigate a dusty \HIIt\ region surrounding a young
stellar population including binary stars. The spectrum of the adopted
population is obtained using \mbox{
  \texttt{pyBPASS}}\footnote{\url{https://gitlab.mpcdf.mpg.de/mglatzle/pybpass}}
and the BPASS v2.2.1 \citep{stanwayReevaluatingOldStellar2018} spectral
database. A stellar metallicity of $0.1Z_\odot$, a Chabrier initial mass
function \citep{chabrierGalacticStellarSubstellar2003} with an upper limit of
\SI{300}{M_\odot}, a stellar age of \SI{5}{Myr}, and a mass of \SI{e6}{M_\odot}
are chosen. The resulting spectrum is down-sampled to 100 bins for use in
\crash, and it extends into the soft X-ray regime ($h\nu \sim \SI{5e2}{eV}$),
which we can consistently treat, including secondary ionization processes. The
source has an ionizing emissivity $Q_0=\SI{2.75e52}{s^{-1}}$, and is placed in
the corner of a $(\SI{60}{pc})^3$ box containing a homogeneous, cosmological
H-He mixture at a density of \SI{10}{cm^{-3}} and dust at a GDR of
\num{50}. These ISM conditions are motivated by average values found around
young stellar populations, resolved in simulated dusty galaxies evolving in the
epoch of reionization ($z \gtrsim 6$) in
\cite{grazianiAssemblyDustyGalaxies2020}.

\begin{figure}
  \centering
  \includegraphics[width=\linewidth]{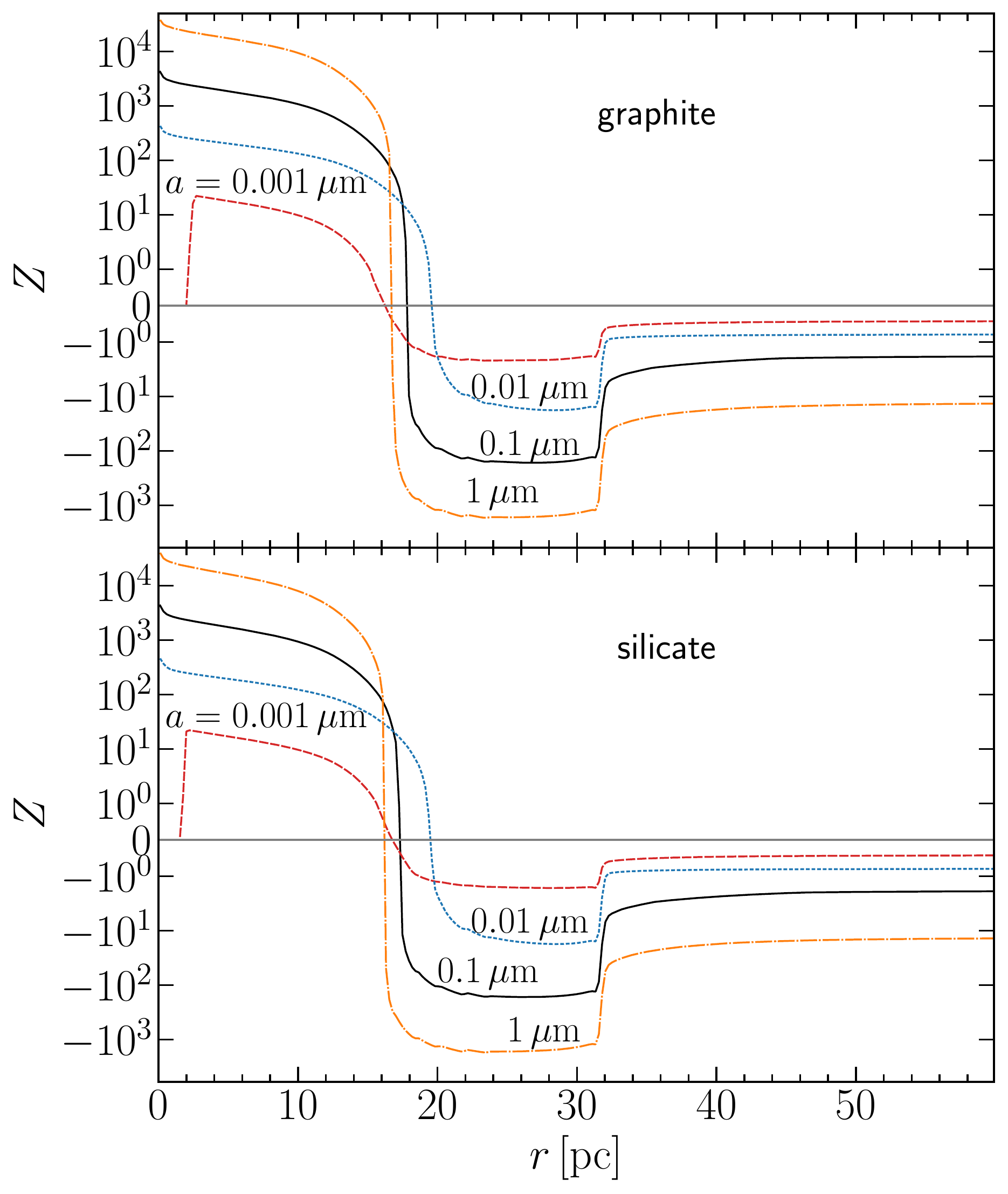}
  \caption{Like Fig.~\ref{fig:BB_Z_all}, but for a source with the spectrum of
    a young stellar population as provided by BPASS. See text for details.}
  \label{fig:BPASS_Z_all}
\end{figure}

Fig.~\ref{fig:BPASS_Z_all} shows the resulting radial profiles of the charges
of grains of different sizes (lines) and materials (panels) at equilibrium. The
qualitative behaviour is similar to that seen in Fig.~\ref{fig:BB_Z_all},
however, the harder spectrum results in higher charge values close to the
source. In the $\sim\SI{2}{pc}$ region closest to the source, the Coulomb
explosion limit is reached for the smallest (\SI{0.001}{\micro m}) grains of
both compositions. Our numerical scheme does not account for grain destruction
yet, so it simply sets the grain charge to zero in this region in order to
clearly identify the possibly affected domain.As discussed in the previous
section, far from the source the value of the negative charge is clearly
affected by the contribution of $J_\mrm{ion}$, as we also verified by comparing
with a similar run disabling its contribution. Differently from the black body
case, on the other hand, the presence of a soft X-ray flux at a large distance
helps stabilizing the computation and removes any numerical artefacts.

The results presented here clearly show the potential of our implementation to
investigate grain charging and some of its consequences as discussed in the
introduction to this chapter. The case of high energy radiation causing partial
gas ionization and thus making available electrons to negatively charge grains
is particularly interesting, as this could be conducive to the growth of
grains. Naturally, a closer investigation is required, taking into account
diffuse matter and source distributions representative of the ISM, as well as
further radiation bands, which could increase the photo-electric charging rate
and drive grain charges towards more positive values. A first step in the
direction of a more realistic model is taken in the next section.

\subsection{Grain charging in a star forming region}
\label{sec:SF-region}
In this section we investigate the effects of radiative transfer on the charge
distribution of cosmic dust grains present in a star forming region of a galaxy
evolving in the epoch of Reionization. To this aim, we set up a more realistic
radiative transfer configuration which extends the previous cases by including
non-H-ionizing frequencies down to a photon energy of \SI{6}{eV} and by
mimicking a realistic gas distribution embedding the sources of radiation in
high density, dust enriched clouds.

\subsubsection{Initial conditions and cloud modelling}
\label{sec:SF-ICs}
The star forming region under investigation is extracted from a galaxy present
in a novel \texttt{dustyGadget} simulation
\citep{grazianiAssemblyDustyGalaxies2020}, which extends the rich chemical
network of \texttt{Gadget} (see \citealt{maioMetalMoleculeCooling2007} and
references therein) by self-consistently accounting for the production of dust
grains by evolving stars \citep{Mancini2015, ginolfiWhereDoesGalactic2018,
  marassiSupernovaDustYields2019} and their successive evolution in the
galactic ISM \citep{hirashitaDustGrowthInterstellar2012}. More specifically,
the adopted simulation is performed on a cubic cosmological volume of
$\sim\SI{70}{cMpc}$ (comoving \si{Mpc}) side length with a gas/dark matter
particle resolution of $\sim\num{8.2e6}/\SI{4.4e7}{M_\odot}$ using the feedback
model introduced in \citet{grazianiAssemblyDustyGalaxies2020}, which is proven
to guarantee a reliable estimate of the total dust mass present in the
simulated galaxies. We note that although this large-scale simulation does not
resolve the detailed structure of star forming regions (see below on how we
address this issue), it nevertheless provides a reliable estimate of the amount
of gas and dust, as well as the stellar content of such regions.

\begin{figure}
  \centering
  \includegraphics[width=\linewidth]{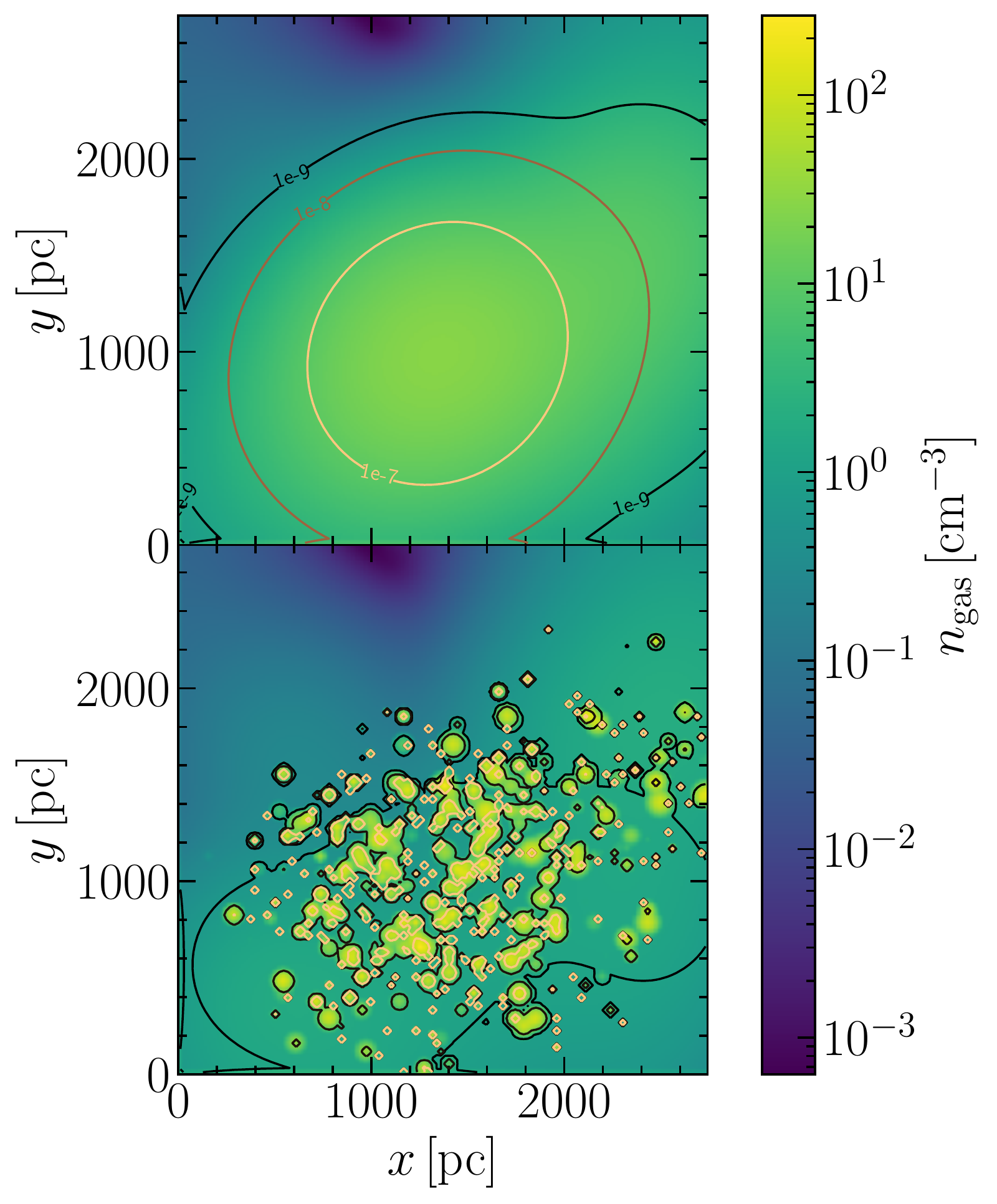}
  \caption{Slice cut of the gas number density distribution in the star forming
    region selected to compute grain charging, before (top) and after (bottom)
    remodelling of the clouds (see text for more details). The corresponding
    dust number density distributions [\si{cm^{-3}}] are shown using contour
    lines.}
  \label{fig:SFRegion}
\end{figure}

Here we select one of the most massive objects present in the cosmological
volume at $z\sim 6.69$ and extract a cubic, isolated galactic region of
\SI{2.8}{pkpc}/side including \num{90} stellar particles with masses around
\SI{2}{M_\odot} and ages varying from \SI{e6}{yr} to \SI{3e8}{yr}. Their
initial metallicity is spread over a range from \SI{5e-4}{Z_\odot} to
\SI{0.1}{Z_\odot}. Their spectra in the range \SIrange{6}{500}{eV} are obtained
from BPASS\footnote{Note that we do not model nebular emission lines in this
  work and that the initial stellar metallicity range supported by BPASS is
  \SIrange{5e-4}{2}{Z_\odot}.} as in \S~\ref{sec:BPASS-source}, resulting in a
spread in ionizing emissivity from \SI{7e47}{s^{-1}} to \SI{2e53}{s^{-1}}. Gas
and dust particles present in this region are projected onto a Cartesian grid
of N$_\mathrm{c} = 128$ cells/side (i.e. the maximum allowed by the
hydrodynamic simulation and equivalent to a spatial resolution of
\SI{21.9}{pc}), to create the necessary initial conditions for a \crash\ run
(see \S~\ref{sec:CRASH} or refer to \citetalias{Glatzle2019} for more details).

Fig.~\ref{fig:SFRegion} (top panel) shows the resulting spatial distribution of
$n_\mrm{gas}$ in a slice cut of the volume selected to contain the brightest
star. At the centre of the image, the gas number density, shown as colour
palette, is $\sim \SI{e2}{cm^{-3}}$ and it rapidly drops by three orders of
magnitude towards the cube edges. The contour lines show the total grain number
density corresponding to the adopted extended MRN size distribution. As
already mentioned above, the mass resolution of the simulation is inadequate to
resolve the clumpy gas structure typical of observed star forming regions. The
total gas and dust mass present in the volume guarantee, on the other hand,
that dust and metal pollution of the region are consistent with all episodes of
star formation and evolution thanks to the accurate dust yields adopted in
\texttt{dustyGadget}.

To address this issue and effectively increase the spatial resolution, we
remodel the distribution of cold gas \citep[see][for a description of the
simulated ISM phases]{grazianiAssemblyDustyGalaxies2020}. The total mass of
cold gas present in the volume is distributed among spherical clouds with a
fixed density of \SI{100}{cm^{-3}} and sizes drawn from a power-law
distribution with index \num{1.1} ranging from \SI{1}{pc} to \SI{100}{pc}. Once
created, the cloud centres are distributed randomly among grid cells, weighting
each cell with its original cold gas content; clouds are allowed to overlap
creating even denser regions. The resulting gas distribution preserves the
original gradient in spherical averages around the cube centre\footnote{We also
  verified that our re-mapping procedure is mass conserving to below
  \SI{1}{\%}.}. The distribution of clouds and diffuse gas is shown in the
bottom panel of Fig.~\ref{fig:SFRegion} (same slice cut as in the top panel),
visually illustrating how the above procedure creates an intricate pattern of
overlapping clouds surrounding or embedding the stellar sources.

The dust associated with the original cold gas content of each cell is finally
distributed among all clouds, weighting with the inverse of the cell-cloud
distance. The resulting dust distribution is illustrated in the panel using
contour lines of the grain number density; note how it statistically correlates
in space with gas clumps by checking the positions of pink iso-contour lines
and yellow regions present in the bottom panel. Also note that all the
iso-contour lines twist together following the cloud clustering; this creates
sudden spatial fluctuations of both gas and dust optical depths along different
lines of sight. Large spatial fluctuations in the radiation field are thus
expected, as a result of radiative transfer effects of photon filtering and
shielding.

The initial temperature of the diffuse gas is taken from the hot phase of the
hydrodynamical simulation, while clouds are assigned the temperature value it
assumes for the cold phase (i.e. $T_\mrm{c} \sim \SI{8e3}{K}$). The initial
charge of the grains is always setup as neutral, and the ionization fractions
for hydrogen and helium are taken from collisional equilibrium calculations.

With the initial conditions described in the above paragraphs, we performed
radiative transfer simulations through gas and dust and computed the grain
charge distribution as described in \S~\ref{sec:chargedCRASH}. The reference
setup propagates \num{e7} photon packets from each source for a simulation
duration of \SI{e6}{yr} and assumes $N_\mrm{d}=200$; its results are verified
to be convergent by running an identical setup with photon sampling increased
by one order of magnitude and testing additional values of $N_\mrm{d}$.

\subsubsection{Gas ionization and temperature}
\label{sec:SF-GasXT}

\begin{figure*}
  \centering
  \includegraphics[width=\linewidth]{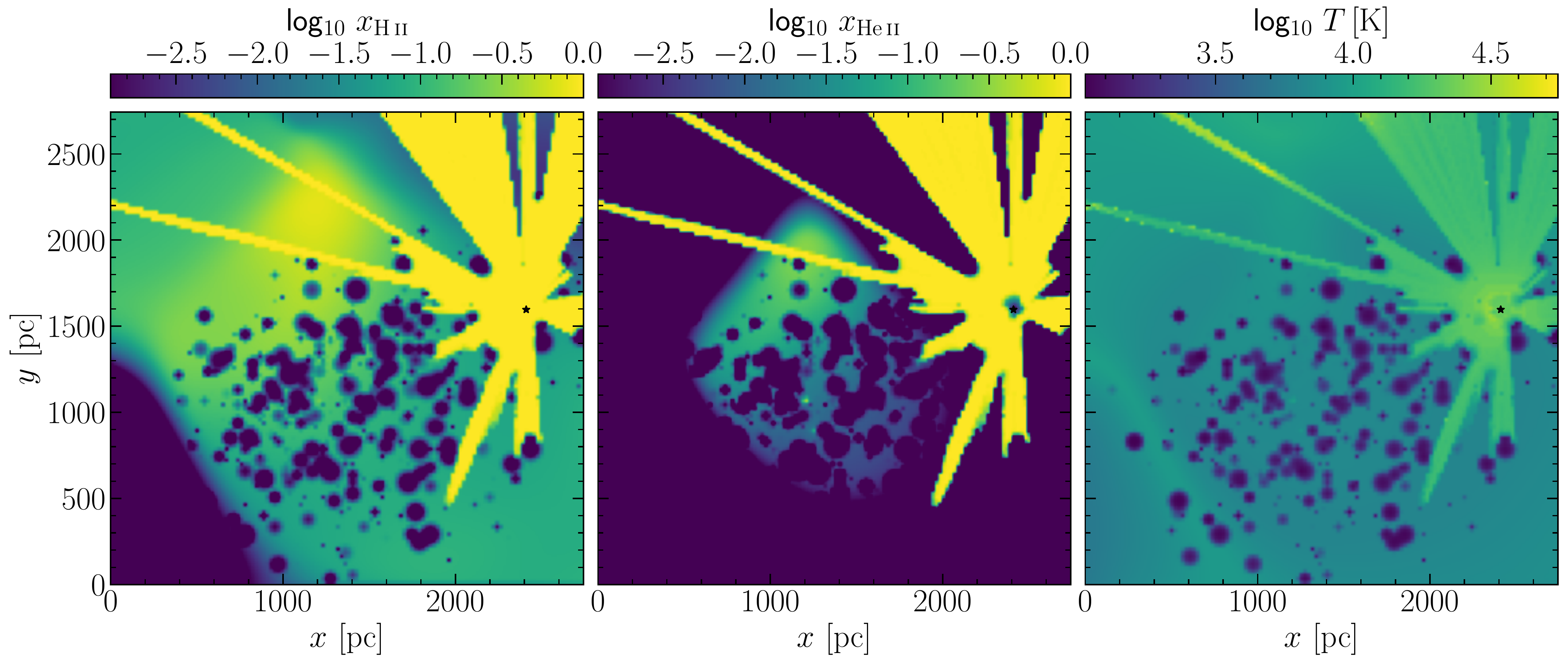}
  \caption{Slice cuts as in Fig.~\ref{fig:SFRegion}, showing hydrogen and
    helium ionization fractions (left and middle panel) and gas temperature
    (right panel) at the end of the radiative transfer simulation.}
  \label{fig:IonsT}
\end{figure*}

Hydrogen ($x_\HII$) and helium ($x_\HeII$) ionization fractions and gas
temperature ($T$) found at the end of the simulation in the usual slice cut are
shown from left to right in Fig.~\ref{fig:IonsT}. The gas close to the
brightest star is fully ionized in both hydrogen and helium (yellow regions in
the left and central panel). Doubly ionized helium (not shown here) is only
present in direct proximity of the brightest star, where $x_\HeII < 1$\,. While
ionizing radiation can easily escape through specific lines of sight and reach
the end of the simulated region (yellow pattern), dusty clouds are very
effective in absorbing these photons and blocking their further propagation. In
fact, even clouds close to the sources can remain significantly neutral in both
H and He. The 3D picture is obviously more complicated as radiation from other
sources, off the slice cut plane, and collisional ionization from hydrodynamic
heating sustain a minimum fraction in ionized hydrogen and singly ionized
helium\footnote{Note that collisional ionization is mainly responsible for the
  differences in the distributions of $x_\HII$ and $x_\HeII$, which are usually
  similar where photoionization by stellar spectra dominates.}. The gas
temperature shown in the right panel is consistent with the ionization
described above: temperatures higher than \SI{e4}{K} are easily found where
ionizing radiation can propagate, while the densest clouds remain at the
initial temperature or even lower (down to \SI{e3}{K}). Finally note that
ionization and temperature maps trace the presence of ionizing radiation only,
while a background of non-ionizing photons is present in the simulated
region. Its behaviour is better reflected by grain charges, as discussed in the
next section.

\subsubsection{Grain charges}
\label{sec:SF-Charges}
\begin{figure}
  \centering
  \includegraphics[width=\linewidth]{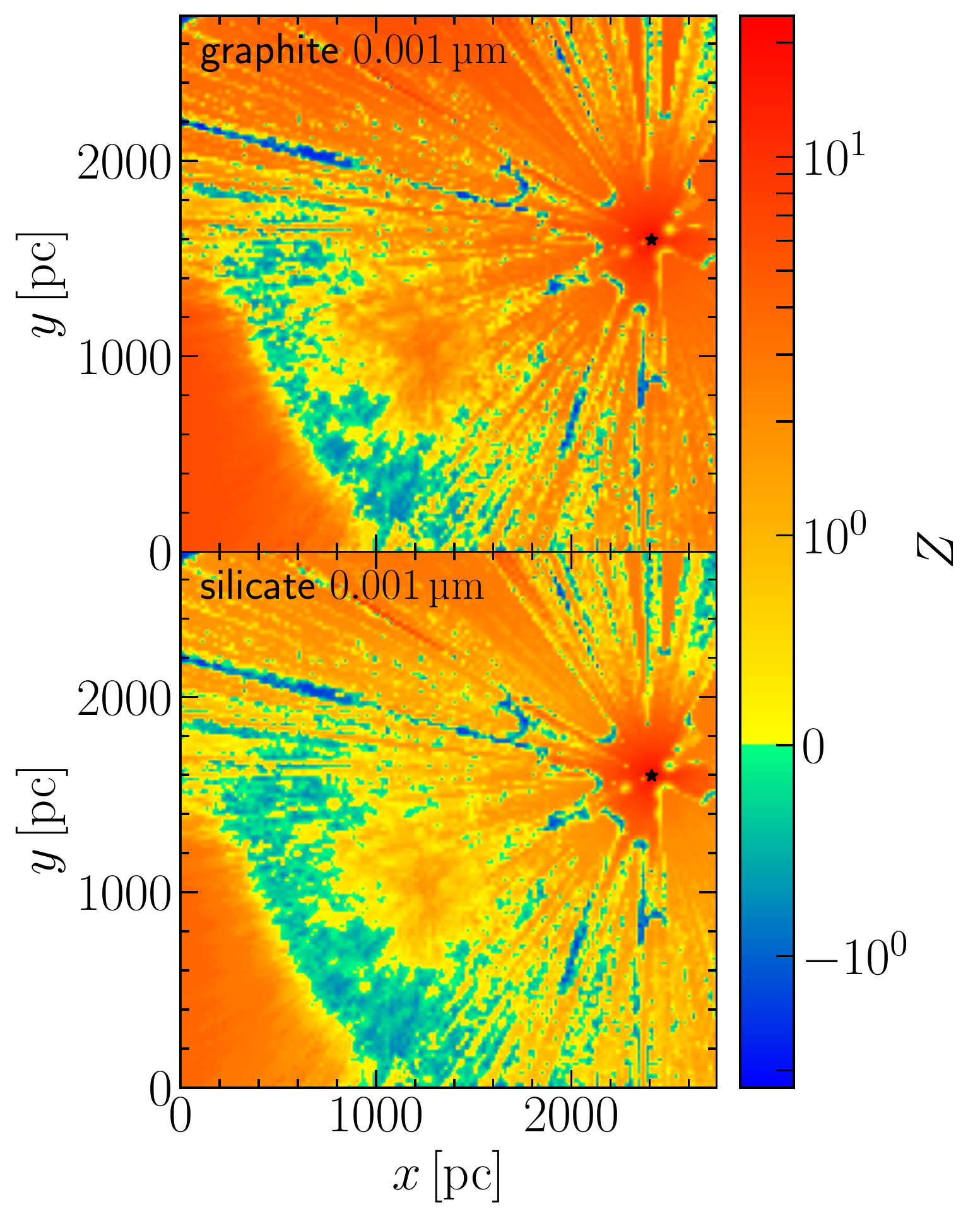}
  \caption{Slice cut of the grain charge spatial distribution in the selected
    star forming region. The charge of graphite grains of reference size
    \SI{e-3}{\micro m} is shown in the top panel, while silicate dust is shown
    in the bottom panel. The colour palette indicates the charge acquired by
    grains at the end of the RT simulation: from yellow to red positive charges
    ($Z \geq 10$), while negative values are represented as gradient toward
    dark blue ($Z \leq -2$).}
  \label{fig:GrainChargeCut}
\end{figure}

The spatial distribution of the grain charges found at the end of the
simulation is shown in Fig.~\ref{fig:GrainChargeCut}. The figure refers to
grains of radius \SI{e-3}{\micro m}; graphite/silicate dust is separately shown
in the top/bottom panel respectively. Positively charged grains are found where
the flux of radiation is high: either coming directly from the brightest source
or the other, off-plane sources. Note that also non-ionizing radiation, the
presence of which is not reflected by the ionization state of the gas, impacts
grain charges. The figure clearly shows that dust grains can get negatively
charged (green/blue regions) if located in clouds sufficiently far from the
brightest source or shielded by other over-densities found along different
lines of sight.

\begin{figure*}
  \centering
  \includegraphics[width=\linewidth]{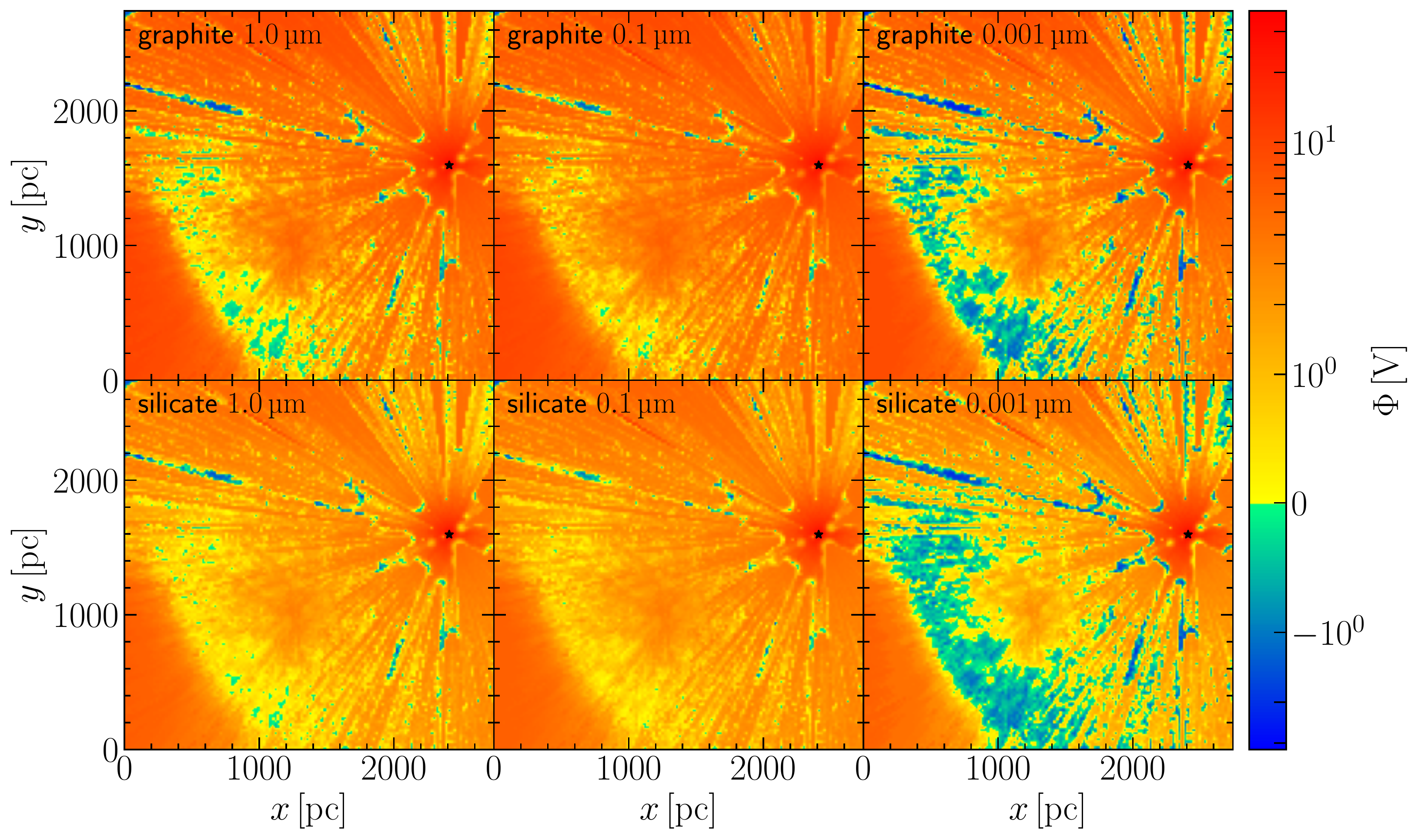}
  \caption{Slice cut of the spatial distribution of grain potential in the
    selected star forming region. Graphite grains are shown in the top panels,
    while silicate dust is shown in the bottom panels. The colour palette
    indicates the potential acquired by grains of different size at the end of
    the RT simulation. From left to right we show grains with \SI{1}{\micro m},
    \SI{0.1}{\micro m} and \SI{e-3}{\micro m}, respectively.}
  \label{fig:GrainSizeCharge}
\end{figure*}

The potential $\Phi = \frac{eZ}{4\pi\epsilon_0 a}$, shown in the composite of
Fig.~\ref{fig:GrainSizeCharge}, is more readily comparable across grains of
different sizes than their charges. By comparing the patterns, from left to
right, it is clear that a negative potential (and thus charge) is favoured in
the smallest grains of both chemical compositions. The graphite panels show a
slightly redder tone than the silicate ones, indicating an overall slightly
more positive potential for graphite grains.

\begin{figure*}
  \centering
  \includegraphics[width=\linewidth]{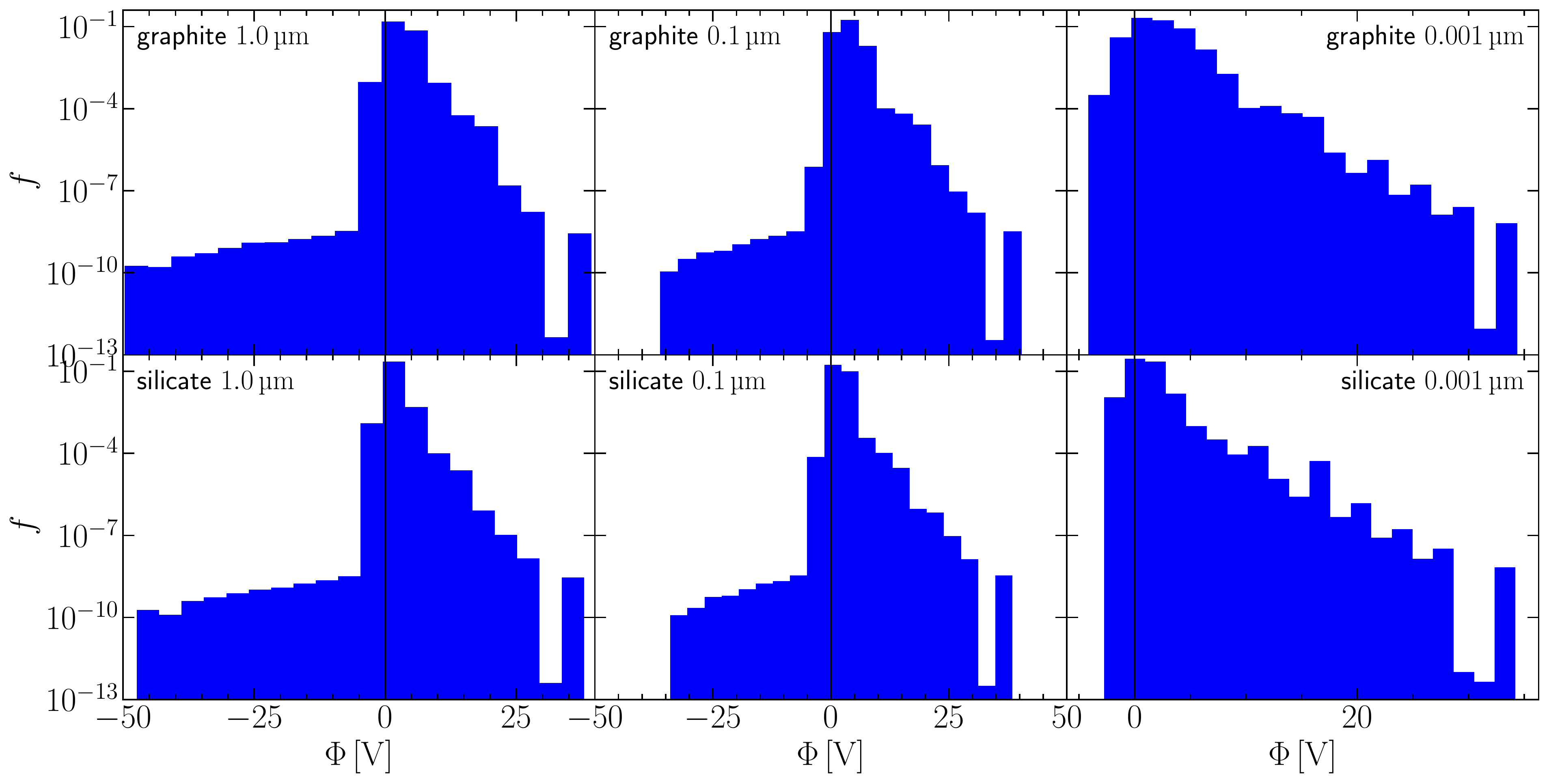}
  \caption{Probability density of the potential, $\Phi$~[V], of graphite (top
    panels) and silicate (bottom) dust grains at the end of the RT
    simulation. From left to right we show grains with \SI{1}{\micro m},
    \SI{0.1}{\micro m} and \SI{e-3}{\micro m}, respectively. }
  \label{fig:GrainSizeChargeHisto}
\end{figure*}

A quantitative picture of the statistics created in the entire volume can be
found in Fig.~\ref{fig:GrainSizeChargeHisto}, which shows the probability
density ($\emph{f}$) of $\Phi$ for the grain sizes and compositions of
Fig.~\ref{fig:GrainSizeCharge}. Note that in the figure the scales and ranges
of $\Phi$ are different across grain sizes. The histograms confirm the
qualitative picture described in the slice cut, with distributions for graphite
grains showing more weight at slightly positive potentials in the range of a
few \si{V}. At the same time, graphite grains reach lower negative potentials
at any size. Negative charges are present for all grain sizes and compositions
although in large grains the probability density is usually very low
($f \le \num{e-5}$); the smallest grains, on the other hand, can reach
potentials as low as \SI{-5}{V} with higher values of $\emph{f}$.

Negative charges are present with different percentage values in each grain
population, strictly depending on their size and chemical composition. For
example, by mass 12\% of the smallest graphite grains (\SI{e-3}{\micro m}) are
negative, while the same value for silicate dust is 21\%. Interestingly, the
equivalent values for the largest grains are instead 4\% and 2\%, respectively.
In this realistic case we note that the lowest values of the potential reached
by intermediate and large grains are determined by $J_\mrm{ion}$, which
prevents values of the potential lower than \SI{-50}{V}/\SI{-30}{V} for grains
with the largest/intermediate radius. The smallest grains of both compositions,
on the other hand, reach their most negative potentials as determined by
auto-ionization.

Finally, it is important to quantify the percentage of negative dust with
respect to the total mass. We find that at the end of the simulation
(i.e. after \SI{1}{Myr}) about 13\% of the total dust mass present in the
simulated volume has a negative charge value; as expected, this is composed by
silicate and graphite grains of radius \SI{e-3}{\micro m} (71\% and 26\%
respectively), with a marginal contribution of all the other larger solid
particles.

Note that a non-H-ionizing galactic background/interstellar radiation field
should be considered here when computing grain charges as we are considering an
inner-galactic region. We defer such a treatment to future work.

\section{Conclusions}
\label{sec:conclusions}
This paper introduces a novel implementation of the grain charging process in
the cosmological radiative transfer code \crash , extending the work of
\citetalias{Glatzle2019}, where the contribution of dust grains was accounted
for only as an additional term to the optical depth of the gas components. In
this new implementation the grain charge is computed consistently with both
radiative transfer effects and the assumed properties of dust grains,
i.e. chemical composition and size distribution. After testing our
implementation on a series of \HIIt\ regions with increasing physical realism,
we studied the case of a complex star forming region extracted from a
\texttt{dustyGadget} simulation \citep{grazianiAssemblyDustyGalaxies2020} and
remodeled its cold gas phase to resolve dusty clouds. At the end of the
simulation (i.e. after \SI{1}{Myr}) we find that $\sim 13$~\% of the total dust
mass remains negatively charged and largely distributed in regions far from the
emitting sources or shielded by dense clouds found along many lines of
sight. In particular:

\begin{itemize}

\item a complex distribution of grain charges is present in the entire volume
  with corresponding potentials varying from \num{-5} to \SI{+50}{V};

\item negative grains account for $\sim 13$~\% of the total grain mass and are
  present in $18\%$ of the simulated volume at the end of the simulation; the
  negative mass is mainly due to silicate and graphite grains of radius
  \SI{e-3}{\micro m} (71\% and 26\% respectively);

\item a clear dependence on grain size emerges from the charges present in the
  simulated volume; a weak dependence on chemical composition is also found as
  graphite grains tend to acquire a slightly more positive potential while at
  the same time reaching lower minimal charges than silicate grains, at fixed
  grain size;

\item grain charge does not directly correlate with gas ionization or
  temperature as it strictly depends on the radiation field, both at ionizing
  and non-ionizing frequencies ($h\nu \geq \SI{6}{eV}$); negative charges seem
  to correlate with the distance from radiation sources, being present in both
  regions far from the sources or with significant shielding, as a result of
  radiative transfer effects.
\end{itemize}

Further investigations will be performed on zoom-in simulations of selected
galaxies simulated with the \texttt{dustyGadget} code to establish the grain
charge distribution in many environments of their resolved, multiphase dusty
ISM.

\section*{Acknowledgements}
We warmly thank the referee for her/his suggestions on our charge
implementation which also significantly improved the quality of the final
results discussed in the present manuscript. LG acknowledges support from the
Amaldi Research Center funded by the MIUR program "Dipartimento di Eccellenza"
(CUP:B81I18001170001). We thank Barbara Ercolano for insightful discussions on
grain charging and Dylan Nelson for sharing his SPH-gridding routines for this
work. Pre- and post-processing as well as plotting were done using the Python
language in its CPython implementation and the Matplotlib
\citep{hunterMatplotlib2DGraphics2007}, NumPy
\citep{harrisArrayProgrammingNumPy2020} and Numba libraries.

\section*{Data Availability}
The data underlying this article will be shared on reasonable request to the
corresponding author.


\bibliographystyle{mnras}
\bibliography{main.bib}



\appendix

\section{New dust cross sections}
\label{sec:new-dust-cross}
Since we encountered problems in reproducing the absorption cross sections of
Silicate-Graphite-PAH mixtures using the Mie code by Wiscombe
\citepalias[][Fig.~3]{Glatzle2019}, after the publication of
\citetalias{Glatzle2019} we adopted the Mie implementation used in
\textsc{Cloudy} \citep[e.g.][]{Ferland2017}, which provides less functionality
but is also less complex. Conveniently, it applies anomalous diffraction theory
in the appropriate limits \citepalias[][App.~B]{Glatzle2019}. We extracted the
relevant portions of source code and combined them into an independent C/C++
library. Moreover, we wrote a CPython extension with a NumPy universal
function\footnote{\url{https://numpy.org/doc/stable/reference/ufuncs.html}}
wrapper for the main routine, such that, after compilation, it can easily be
called from Python with full support of NumPy array arguments. The code has
been
published\footnote{\url{https://gitlab.mpcdf.mpg.de/mglatzle/cosmic_dustbox}}.

Using this new library in combination with dielectric functions and PAH cross
sections as described in \citetalias[][App.~B]{Glatzle2019}, we find far better
agreement with the published cross sections at all wavelengths.

\section{Photon counting inconsistency}
\label{sec:dust-fix}
\begin{figure}
  \centering
  \includegraphics[width=\linewidth]{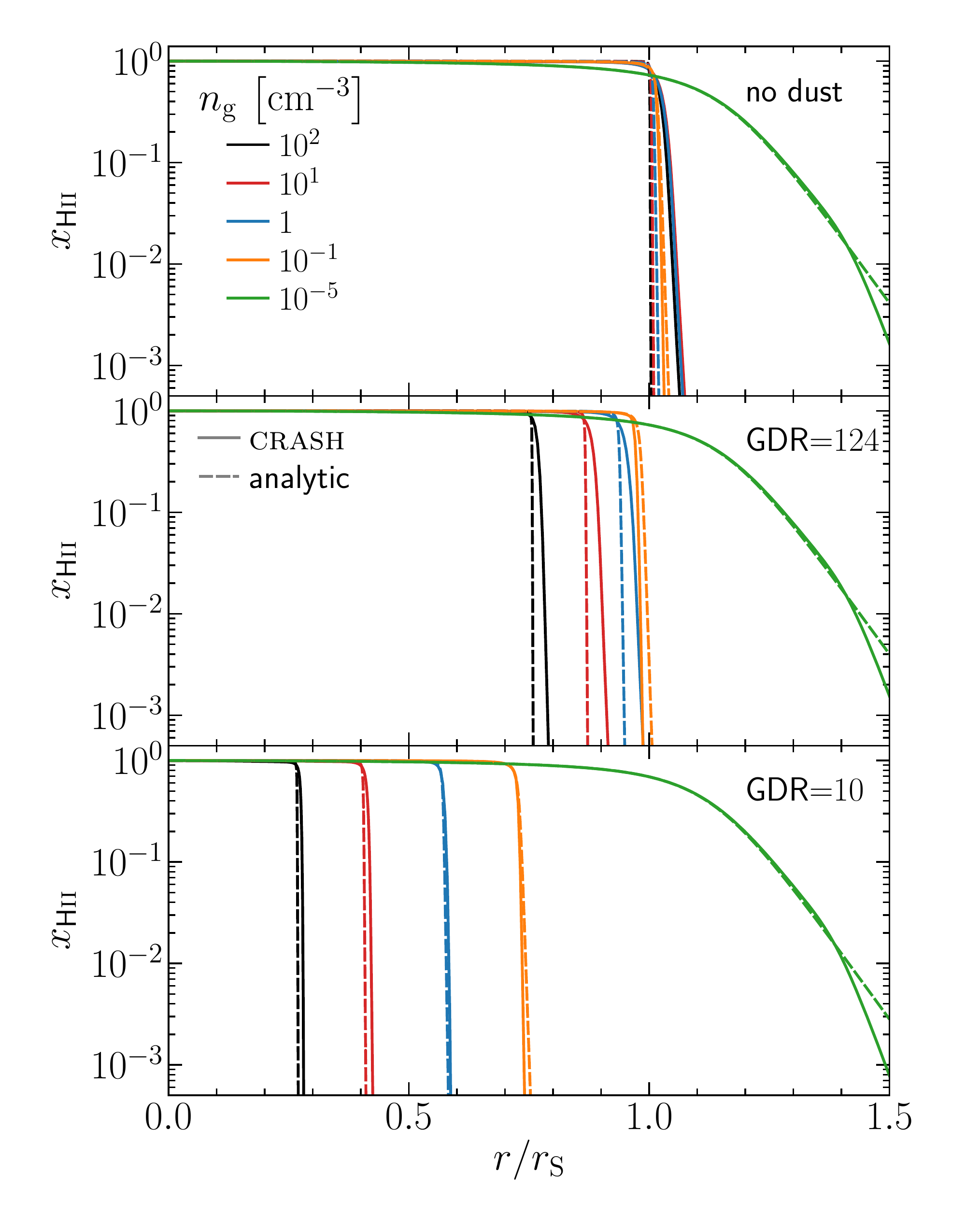}
  \caption{Hydrogen ionization fraction as a function of distance from the
    ionizing source normalized to the Str\"omgren radius for different gas
    densities (line colours) and gas to dust mass ratios (panels). The analytic
    solution \citep[cf.][]{Raga2015} is shown in dashed and the corresponding
    \crash\ results are shown in solid. See text for details.}
  \label{fig:raga}
\end{figure}

We discovered an inconsistency in how ionizing photons absorbed by dust were
counted in the \crash\ version used in \citetalias{Glatzle2019}. This
inconsistency resulted in too many photons being removed from photon packets
and thus an overprediction of dust absorption. The effect is strongest in the
regime of high gas densities and low GDRs, in which we reported disagreement
with the analytic solution in App.~A of \citetalias{Glatzle2019} and which we,
in light of our interest in more tenuous environments and this disagreement,
avoided in the other simulations presented in \citetalias{Glatzle2019}. The
issue has been fixed and we now find much better agreement with the analytic
solution in a wide parameter range. This is illustrated in Fig.~\ref{fig:raga},
where we show the hydrogen ionization fraction as a function of distance from
the ionizing source normalized to the Str\"omgren radius, $r_{\rm S}$, for
different gas densities (lines) and GDRs (panels). The analytic solution,
computed as in \cite{Raga2015}, is shown in dashed and the corresponding
\crash\ results are shown in solid. The simulations were set up as described in
App.~A of \citetalias{Glatzle2019}.

We consider the agreement to be excellent, taking into account the stark
differences in the two approaches. For possible origins of the remaining small
deviations, see the discussion in App.~A of \citetalias{Glatzle2019}.

\section{Computation of photo-electric yields}
\label{sec:photo-el-yields}
Following \citetalias{Weingartner2001b} and \citetalias{Weingartner2006}, we
implemented a Python framework to compute the photo-electric yield for a grain
of a given species (carbonaceous or silicate), size and charge at a given
photon energy $h\nu$. Here we provide some details on our implementation, an
experimental version of which has been made publicly
available\footnote{\url{https://gitlab.mpcdf.mpg.de/mglatzle/cosmic_dustbox}}.

In the computation of the photon attenuation length $l_\mathrm{a}$ (see
\citetalias{Weingartner2001b}), we use the same dielectric functions as in
App.~\ref{sec:new-dust-cross} \citepalias[][App.~B]{Glatzle2019} to be
consistent.

In \S~5, \citetalias{Weingartner2006} state first that they use
$E_{\mathrm{A},i,s,j}$ when evaluating the electron escape length
$l_\mathrm{e}$ to compute $y_{0,\mathrm{A}}$ (Eq.~(16)). In the next paragraph,
however, they state that they use $\Theta_{\mathrm{A};i,s}$ when evaluating
$y_{0,\mathrm{A}}$ and $\Theta_{\mathrm{A};i,s}$ also appears as an argument to
$y_{0,\mathrm{A}}$ in Eq.~(16). It is not clear to us where
$\Theta_{\mathrm{A};i,s}$ could be required to evaluate Eq.~(16) other than as
a replacement for $E_{\mathrm{A},i,s,j}$ as the argument to
$l_\mathrm{e}$. This would, however, effectively introduce a direct dependency
of the initial Auger electron energy on the energy of the photon that created
the inner shell vacancy, the filling of which initiates the Auger emission
process. We do, therefore, not use $\Theta_{\mathrm{A};i,s}$ in Eq.~(16) (or
anywhere in our calculation, for that matter) and evaluate $l_\mathrm{e}$ using
$E_{\mathrm{A},i,s,j}$.

When computing secondary electron yields (\S~6), \citetalias{Weingartner2006}
adopt bulk values for the initial energy of primary electrons ejected from band
structure and inner shells (e.g. $E_{e,\mathrm{band}}=h\nu-W$) in the
evaluation of Eq.~(18). For consistency with the computation of primary yields,
we set ${E_{e,k}=\Theta_k}$. This has no discernible effect on the results (see
next paragraph) and is certainly negligible in light of all other uncertainties
in the properties of grain materials.

In the top panel of Fig.~\ref{fig:pe-yield-sizes}, we reproduce Fig.~4 from
\citetalias{Weingartner2006} with very good agreement. The only difference
discernible to the eye is the first peak in the $a=\SI{0.001}{\micro\meter}$
curve at ${\sim\SI{10}{eV}}$. While our peak reaches~$\sim 0.9$, their peak
reaches only $\sim 0.7$\,. This is owed to a small bug in the code that was
used to plot Fig.~4 of \citetalias{Weingartner2006} (J.~Weingartner, private
communication).

\section{Comparison with analytic solution}
\label{sec:comp-with-analyt}
\begin{figure}
  \centering
  \includegraphics[width=\linewidth]{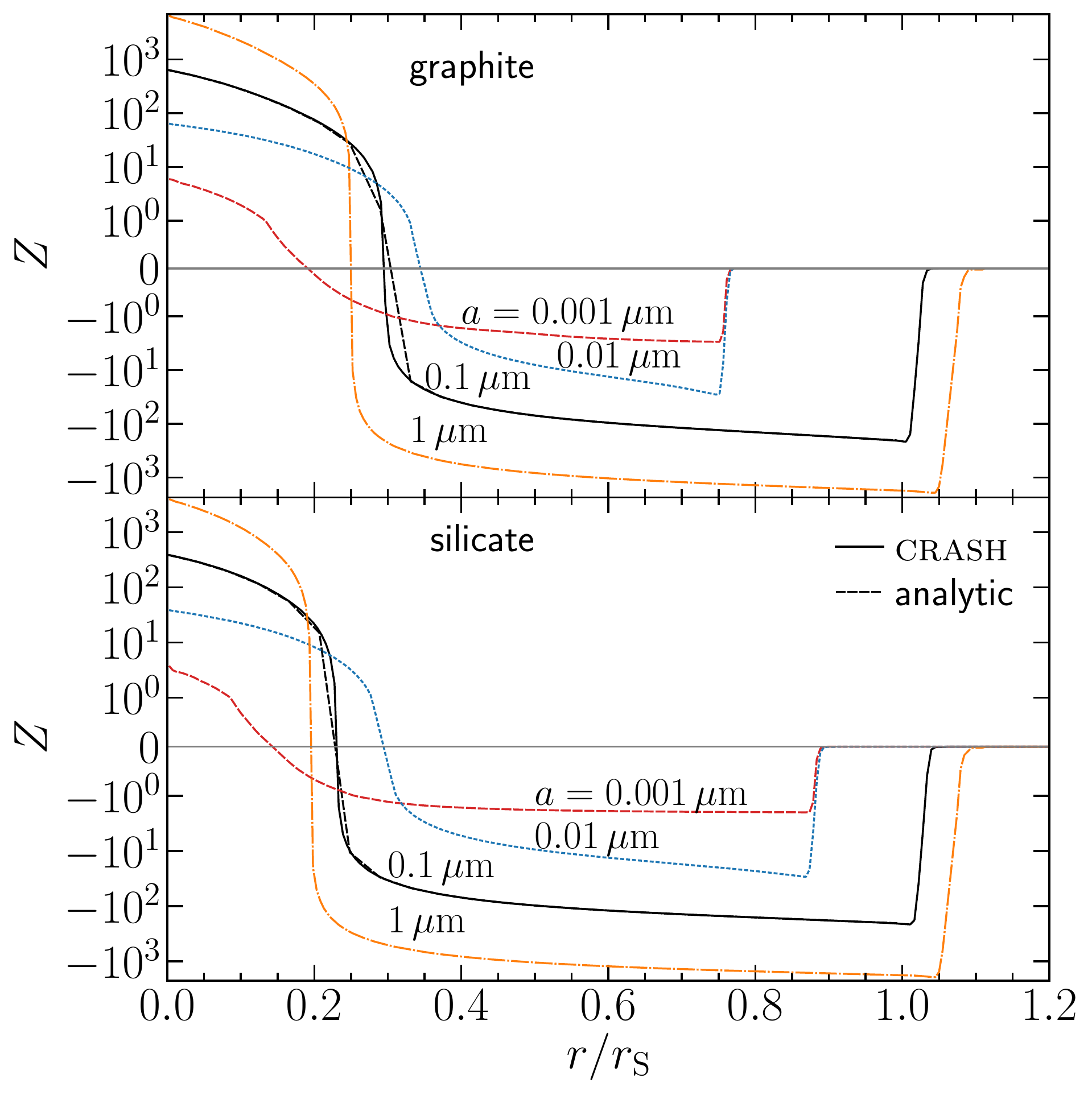}
  \caption{Grain charge at equilibrium as a function of distance from an
    ionizing source emitting at \SI{13.6}{eV} normalized to the Str\"omgren
    radius for simulations including grains of different sizes (line styles)
    and chemical compositions (panels). The gas number density is
    \SI{1}{cm^{-3}} and GDR$=\num{124}$. For $a=\SI{0.1}{\micro m}$, we also
    show an analytic estimate (dashed black) for comparison to \crash\ (solid
    black). The horizontal line is drawn to guide the eye. See text for
    details.}
  \label{fig:raga-Z}
\end{figure}
The analytic solution to the idealized, equilibrium dusty \HIIt\ region problem
used in App.~\ref{sec:dust-fix} provides the hydrogen ionization fraction
$x_\HII$ and the ionizing photon flux as a function of radius from the
source. While it is possible to modify it to account for dust charging
\citep[cf.][\S~7.7]{Osterbrock2006}, here we take a simpler approach. For
typical GDRs ($\sim\num{100}$), if hydrogen is highly ionized, dust charging
will not appreciably change the number density of free electrons. This is
clearly the case close to the source, while it is not necessarily true towards
the ionization front. We therefore compute the grain charge using $x_\HII$ and
the flux as given by the analytic solution, starting from the source and moving
outward until we reach a point where the thus predicted charge value would
result in a change of the free electron number density of more than one
percent. At this point we stop the computation.

Fig.~\ref{fig:raga-Z} shows spherical averages of the grain charge $Z$ for
\crash\ simulations featuring dust populations consisting of grains of a fixed
radius (lines) and composition (panels). The simulation setup is identical to
that of the simulations shown in Fig.~\ref{fig:raga} with
$n_\mrm{g}=\SI{1}{cm^{-3}}$ and $\text{GDR}=124$, except for the fact that all
dust mass is comprised of identical grains. Initially, the grain charge is set
to zero and it is then evolved until equilibrium is reached. We show the
analytic solution as discussed above only for $a=\SI{0.1}{\micro m}$ so as to
not overcrowd the plot.

Close to the source, the charge values are positive due to the high ionizing
flux. At some distance, a transition to negative charge values takes place as
free electrons from hydrogen ionization continue to be highly abundant, while
the flux strongly decreases. The transition is sharper for larger grains. The
charge then remains negative up to the ionization front where it returns to the
initial zero value. The position of the ionization front varies depending on
grain size and composition, since we keep the mass of dust constant and small
grains offer a higher opacity per mass (cf.~Fig.~\ref{fig:sigma-d-sizes}). The
agreement with the analytic solution is excellent up to $\sim r_\mrm{S}$, where
we stop computing it as it is no longer valid. This is also true for the other
grain sizes (not shown).


\bsp	
\label{lastpage}
\end{document}